\documentclass[reprint,prb,longbibliography,superscriptaddress]{revtex4-2}

\usepackage{amssymb, amsmath}
\usepackage[english]{babel}
\usepackage{bm}
\usepackage{graphicx}
\usepackage[utf8]{inputenc}
\usepackage{bbold}

\usepackage[dvipsnames]{xcolor}

\newcommand{\commenttoggle}[1]{}

\newcommand{\red}[1]{{ #1}}

\mathchardef\myminus="2D

\usepackage{letltxmacro}
\LetLtxMacro{\ORIGselectlanguage}{\selectlanguage}
\makeatletter
\DeclareRobustCommand{\selectlanguage}[1]{%
  \@ifundefined{alias@\string#1}
    {\ORIGselectlanguage{#1}}
    {\begingroup\edef\x{\endgroup
       \noexpand\ORIGselectlanguage{\@nameuse{alias@#1}}}\x}%
}
\newcommand{\definelanguagealias}[2]{%
  \@namedef{alias@#1}{#2}%
}

\makeatother

\definelanguagealias{en}{english}
\definelanguagealias{EN}{english}
\definelanguagealias{eng}{english}
\definelanguagealias{de}{ngerman}

\begin{document}
\author{Marten Richter}
\email[]{marten.richter@tu-berlin.de}

\affiliation{Institut für Theoretische Physik, Nichtlineare Optik und
Quantenelektronik, Technische Universität Berlin, Hardenbergstr. 36, EW 7-1, 10623
Berlin, Germany}

\title{Theory of interlayer exciton dynamics in 2D TMDCs Heterolayers under the influence of strain reconstruction and disorder}



\begin{abstract}
Monolayers of transition metal dichalcogenides (TMDC) became one of the most studied nanostructures in the last decade. Combining two different TMDC monolayers results in a heterostructure whose properties can be individually tuned by the twist angle between the lattices of the two van-der-Waals layers and the relative placement of the layers, leading to Moiré cells. For small twist angles, lattice reconstruction leads to strong strain fields in the Moiré cells. In this paper, we combine an existing theory for lattice reconstruction with a quantum dynamic theory for interlayer excitons and their dynamics due to exciton phonon scattering using a polaron transformation. The exciton theory is formulated in real space instead of the commonly used quasi-momentum space to account for imperfections in the heterolayer breaking lattice translational symmetry. We can analyze the structure of the localized and delocalized exciton states and their exciton-phonon scattering rates for single phonon processes using Born-Markov approximation and multi-phonon processes using a polaron transformation. Furthermore, linear optical spectra and exciton relaxation Green functions are calculated and discussed.
\red{A P-stacked MoSe$_2$/WSe$_2$ heterolayer is used as an illustrative example. It shows excitons localized in the potential generated through the Moiré-pattern and strain and a delocalized continuum. The exciton-phonon relaxation times vary depending on the strain and range from sub-pico seconds up to nanoseconds.}
\end{abstract}


\date{\today}
\maketitle

After the invention and widespread research of monolayer graphene \cite{doi:10.1126/science.1102896,geim2009graphene,PhysRevB.83.153410,malic2011microscopic,PhysRevB.105.085404} with its special linear band structure around the K and K' valleys \cite{sprinkle2009first,PhysRevB.78.075435}, other monolayer materials with band gaps and thus suitable for optoelectronic applications came into the research focus \cite{PhysRevB.93.125428,vannucci2020anisotropic,novoselov2005two}. One interesting aspect is the optical selectability of different K and K' valleys with polarisation \cite{PhysRevLett.108.196802,mak2012control,PhysRevB.86.081301}. The most prominent group of materials are transition metal dichalcogenides (TMDC) monolayers, which dominated a large part of semiconductor research in physics for the last decade \cite{groenendijk2014photovoltaic,korn2011low,volzer2021fluence,berghauser2014analytical,PhysRevB.101.155304,gao2016valley,moody2015intrinsic,wierzbowski2017direct,lohof2018prospects}. One major topic for understanding the optical properties of TMDCs is devoted to understanding optically induced electron-hole complexes predominantly excitons, but also trions and biexcitons \cite{PhysRevB.93.041401,PhysRevB.101.075302,PhysRevB.99.241301, hao2017neutral,PhysRevB.96.075431,doi:10.1021/acs.nanolett.8b00840,druppel2017diversity,PhysRevLett.131.146201,PhysRevB.108.L121102,denning2022bichromatic}.
In this area, monolayer TMDCs as two-dimensional systems are analog to quantum wells in earlier two-dimensional semiconductor nanostructures.
For the earlier quantum wells, heterostructures were formed by combining multiple quantum wells \cite{PhysRevLett.92.067402, stroucken1996coherent,dang2010influence,PhysRevLett.85.2002,ZIMMERMANN200389,kira1999quantum,chen2007effects,haug2009quantum}. The analog for TMDCs monolayers are the hetero-layers formed from different monolayers such as a MoSe$_2$ monolayer and a WSe$_2$ monolayer.
One significant difference to quantum wells is that the different monolayers are weakly bound by the van-der-Waals force instead of the strong covalent bounding in the quantum well case. 

\begin{figure}[b]
    \centering
    \includegraphics[width=8.6cm]{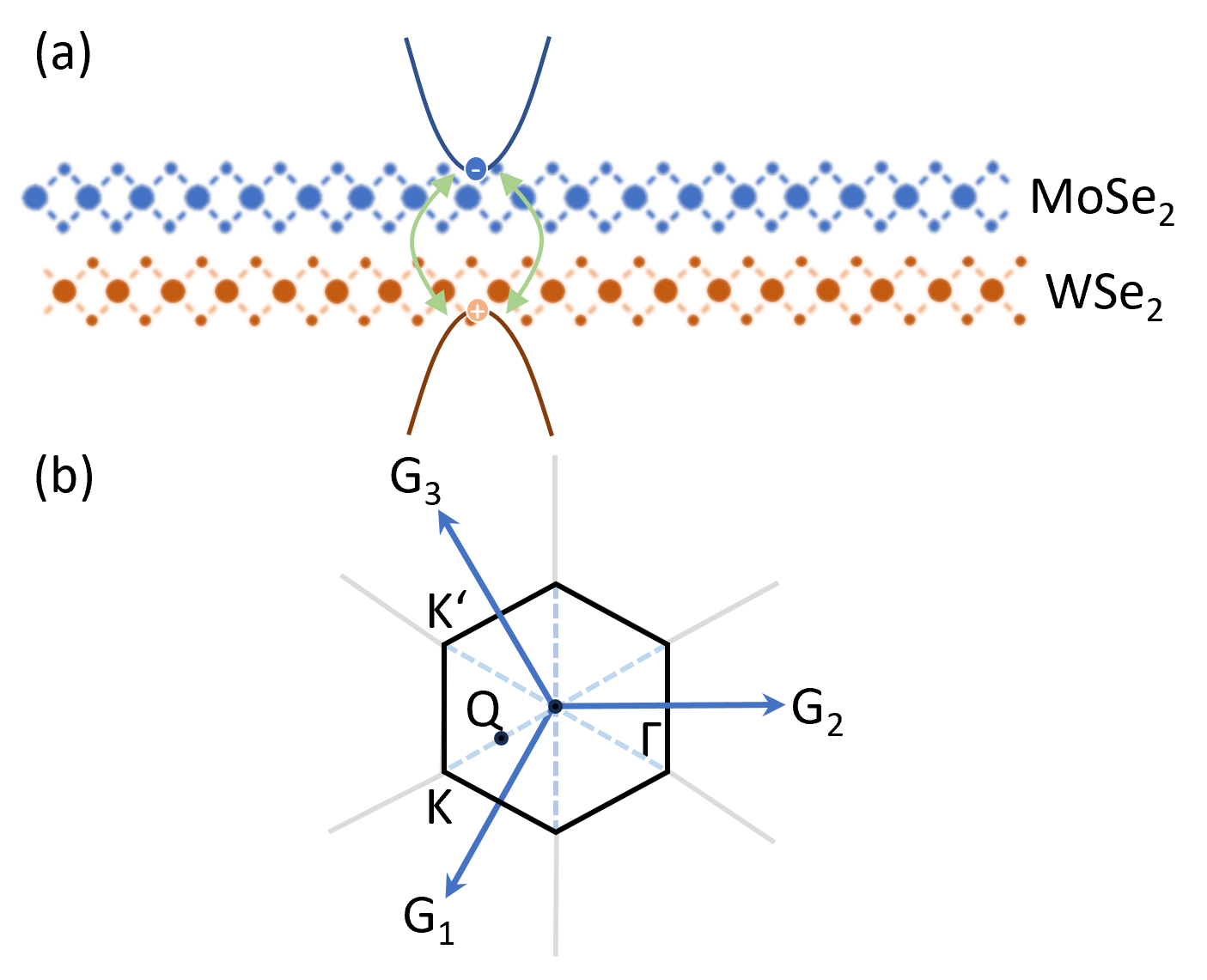}
    \caption{\red{(a)} Illustration of the interlayer exciton formed between a hole in WSe$_2$ and an electron in MoSe$_2$, \red{(b) Illustration of the monolayer Brioullin zone with the special points K, K', Q and $\Gamma$ and the reciprocal lattice vectors $G_i$.}}
    \label{illustrationinterlayerX}
\end{figure}
The strong covalent bounds in the quantum well case limited the realized heterostructures to materials with very close lattice constants. Also, the orientation of unit cells of the materials to each other is set by the covalent binding. This is not the case for the van-der-Waals-bound materials. The two layers can be moved and rotated relatively freely to each other. The lattice mismatch and relative positioning lead to the formation of Moiré supercells \cite{PhysRevB.102.195403}, whose size is determined by the lattice mismatch and the local stacking angle $\theta$. These layers are often embedded inside hBN layers to stabilize the structure mechanically.
Recently, it has been suggested by \cite{PhysRevLett.124.206101,arxiv.2202.11139,enaldiev2021piezoelectric,kaliteevski2023twirling,soltero2023competition,enaldiev2022self} that mechanical interactions lead to mechanical reconstruction, changing the strain field in the unit cells of the Moiré pattern substantially. 
For quantum wells also in almost perfectly grown samples, certain fluctuations in alloys or through interface roughness still lead to a small degree of disorder. Conversely, the strain, through incomplete relaxation or reconstruction or at borders or cracks, may serve as a source of disorder for the TMDC heterostructure \footnote{Note, we consider very little disorder, but even the best experimental procedure never reaches the idealized theoretical models, so studying the disorder's influence is worthwhile.}.

To achieve this, we combined the available theory for strain reconstruction \cite{PhysRevLett.124.206101,arxiv.2202.11139,kaliteevski2023twirling} in Sec. \ref{sec:strain_recon} together, with exciton theory with the disorder in real space adapted from quantum wells \cite{ZIMMERMANN200389,PhysRevB.95.235307} in Section \ref{sec:interlayer_exc_states}. The exciton theory also includes the electron and hole single-particle potentials originating from strain \cite{Khatibi_2019,PhysRevB.97.035306,tran2019evidence}. In Sec. \ref{sec:compdomianexcstates}, the overall structure of the interlayer exciton states is analyzed for one first example.
In the following section, exciton-phonon (with and without polaron transformation) and exciton-photon coupling and rates are derived and analyzed for the example structure. Finally, in Sec. \ref{sec:eqmonumres}, linear optical spectra, as well as exciton relaxation, are analyzed.
Overall, a theoretical framework is set up, which will be used to analyze different TMDC (interlayer) excitons under the influence of strain, including disorder or other imperfections such as cracks, holes, or borders.

As an example to illustrate the theory, we focus on interlayer exciton in a MoSe$_2$/WSe$_2$-hetero bilayer for one specific configuration
\red{ (cf. Fig. \ref{strainandpots}a)).
The lowest energy difference between the valence and conduction bands is at the K and K' points (cf. Fig. \ref{strainandpots}b)) of the Brioullin zone of the monolayer.
Thus, the lowest energy intralayer exciton states reside at K and K' points.
The interlayer exciton of MoSe$_2$/WSe$_2$-hetero bilayer with the electron in the K valley of the valence band of MoSe$_2$ layer and hole in the K valley of the conduction band of WSe$_2$ layer.
}

\section{Strain reconstruction}\label{sec:strain_recon}
To retrieve the strain field for the subsequent calculations,
we aim to obtain the mechanical strain of the hetero bilayer close to equilibrium with some remaining disorder. This is the first step before a quantum dynamical calculation of the exciton migration. 
For calculating the strain fields, we follow the approach from \cite{PhysRevLett.124.206101} with elements also from \cite{arxiv.2202.11139}.
Generally, we use periodic supercells of a reconstructed Moiré pattern with different sizes and aspect ratios for the computational domain. 
We start from the energy of the lattice containing the intralayer $U(\mathbf{r})$ and interlayer $W_s(\mathbf{r})$ strain contributions \cite{PhysRevLett.124.206101}
\begin{eqnarray}
\mathcal{E}=\int_S\left[U(\mathrm{r})+W_s(\mathrm{r})\right]d^2r.
\end{eqnarray}
\begin{figure}[bth]
    \centering
    \includegraphics[width=8.6cm]{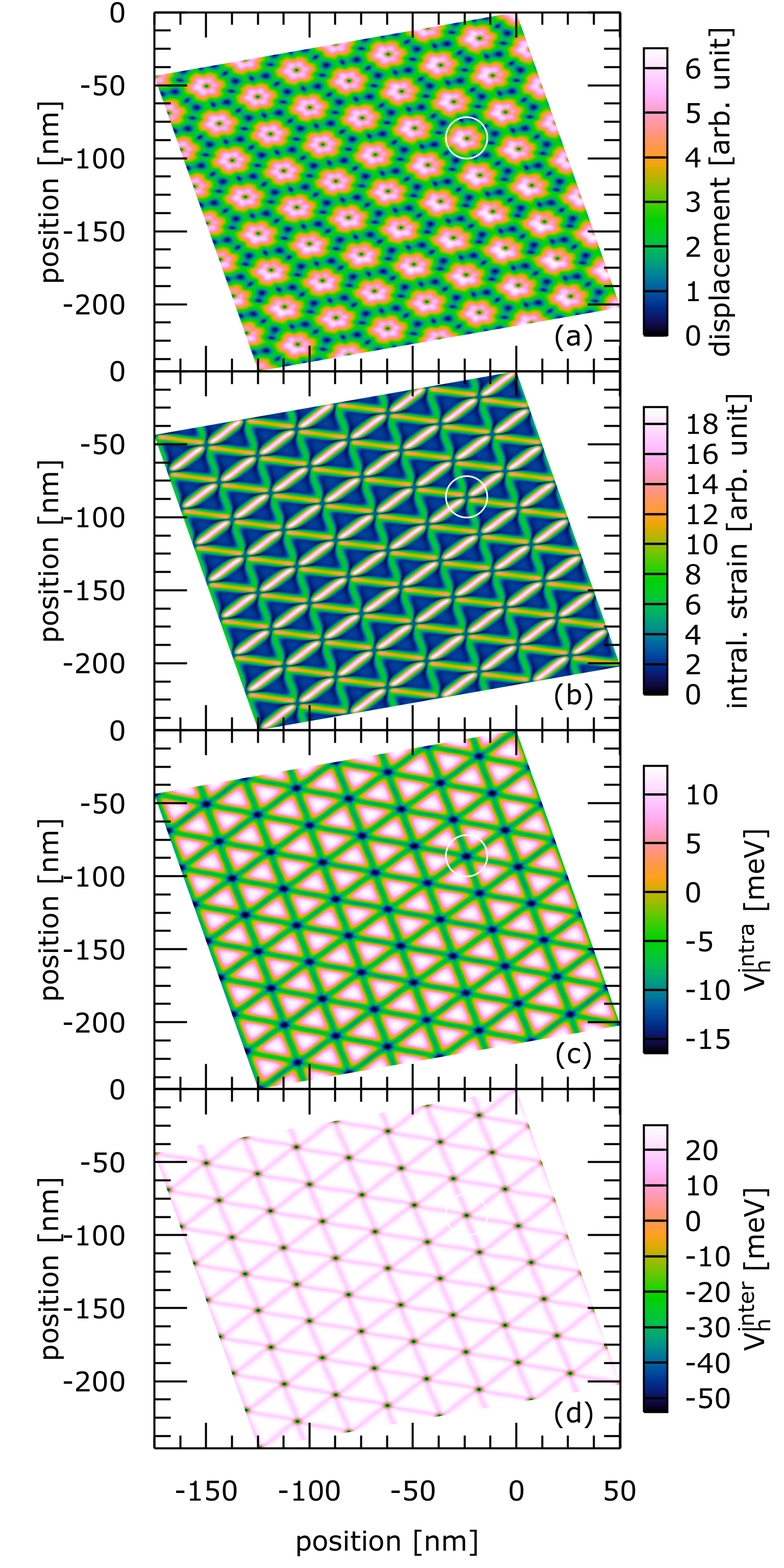}
    \caption{For the WSe$_2$/MoSe$_2$ heterolayer example, (a) the displacement $|\mathrm{u}_{\cdot}|$ of the bottom layer, (b) the absolute intralayer strain of the bottom layer $|\partial u_x/\partial x|+ |\partial u_y/\partial y|$, (c) intralayer hole potential $V_{h}^{intra}(\cdot)$ and (d) interlayer hole potential \red{$\frac12V_{h}^{inter}(\cdot)$}. \red{The white circle marks the position of state $1$ for visual orientation.}}
    \label{strainandpots}
\end{figure}
The intralayer contribution reads \cite{PhysRevLett.124.206101} 
\begin{eqnarray}
U(\mathrm{r})=\sum_{l=t,b} \left[\frac{\lambda_l}{2} (u_{i i}^{(l)})^2+\mu_l(u_{i j}^{(l)})^2\right],
\end{eqnarray}
with the first Lamé coefficient $\lambda_{t/b}$, shear modules $\mu_{t/b}$ and the strain tensors $u_{i j}^{(t/b)}=\frac12(\partial_j u_i^{(t/b)}+\partial_i u^{(t/b)}_j)$ describing local in-place displacement in top or bottom layer respectively  \cite{PhysRevLett.124.206101}. 

\begin{table}
\label{parametertable}
\caption{\label{paratable}Parameters used for the illustrative calculations of a MoSe$_2$/WSe$_2$ hetero bilayer.}
\begin{ruledtabular}
\begin{tabular}{ll}
Mechanical and grid parameters \\
\hline
$a_1$ = 0.329 nm, $a_2$ = 0.328 nm,\\
$d_0$ = 0.69 nm,
$\rho$ = 0.052 nm,\\
$\lambda_{MoSe_2}$ = 263.982  eV/nm$^2$,
$\lambda_{WSe_2}$ = 185.227 eV/nm$^2$,\\
$\mu_{MoSe_2}$ = 309.892 eV/nm$^2$,
$\mu_{WSe_2}$ = 302.212 eV/nm$^2$,\\
$A_1$ = 77621500. eV/nm$^2$,
$A_2$ = 84739. eV/nm$^2$,\\
$\epsilon$ = 189. eV/nm$^4$ &\cite{PhysRevLett.124.206101}\\
\hline
Exciton parameters\\
\hline
$V_{exc}$ = 0.0018 eV (AP stacking) &\cite{PhysRevB.97.035306}\\
$V_{exc}$ = 0.018 eV (P stacking) &\cite{tran2019evidence}\\
$\Delta E^{h}_{gap}$ = 5.3 \red{eV} (WSe$_2$),\\
$\Delta E^{e}_{gap}$ = -6.9 \red{eV} (MoSe$_2$)  &\cite{Khatibi_2019} \\
$m_e$ =  0.80  (MoSe$_2$),
$m_h$ = 0.45    (WSe$_2$)   &\cite{PhysRevB.102.195403}\\
$\epsilon_r$ = 2.4  &\cite{PhysRevB.102.195403} \\
$r_*$ =
    $3.979 ~\mathrm{nm} / \epsilon_r$  (MoSe2),\\
$r_*'$ = $4.511 ~\mathrm{nm} / \epsilon_r$ (WSe2)  &\cite{PhysRevB.102.195403,PhysRevLett.116.086601}\\
$d$ = 0.702 nm              &\cite{PhysRevB.96.075431,PhysRevB.102.195403} \\
\hline
Phonon parameters\\
\hline
$D_{1,ac}^{c}$ = 3.4 eV (MoSe$_2$),
$D_{0,opt}^{c}$ = 58 eV/nm (MoSe$_2$)            &\cite{PhysRevB.90.045422} \\
$\rho$ = 27837133 fs$^2$ eV/nm$^4$ (MoSe$_2$)  &\cite{PhysRevB.98.035302}\\
$v_A$ =    0.0041 nm/fs (MoSe$_2$)  &\cite{PhysRevB.90.045422} \\
 $\omega_{LO,\Gamma}$ = 36.6 meV,
$\omega_{TO,\Gamma}$ = 36.1 meV,\\
$\omega_{A1,\Gamma}$ = 30.1 meV, 
 $\omega_{LO,Q}$ = 37.5 meV,\\
 $\omega_{TO,Q}$ = 36.4 meV,
 $\omega_{A_1,Q}$ = 27.1 meV (MoSe$_2$)  &\cite{PhysRevB.90.045422}  \\
 \hline
 $D_{1,ac}^v$ = 2.1  eV (WSe$_2$),
$D_{0,opt}^v$ = 31  eV /nm  (WSe$_2$),  &\cite{PhysRevB.90.045422}  \\
$\rho$ = 37698719 fs$^2$ eV/nm$^4$ (WSe$_2$)  &\cite{PhysRevB.98.035302} \\
$v_A$ =  0.0033 nm/fs   (WSe$_2$)&\cite{PhysRevB.90.045422}  \\
 $\omega_{LO,\Gamma}$  = 30.8 meV,
$\omega_{TO,\Gamma}$ = 30.5 meV, \\
$\omega_{A1,\Gamma}$ = 30.8 meV, 
 $\omega_{LO,Q}$ = 32.5 meV, \\
 $\omega_{TO,Q}$ = 27.3 meV,
 $\omega_{A_1,Q}$ = 30.4 meV (WSe$_2$) &\cite{PhysRevB.90.045422} 
\end{tabular}
\end{ruledtabular}
\end{table}

The interlayer contribution has the form \cite{PhysRevLett.124.206101,arxiv.2202.11139} $W_s(\mathbf{r})=V_s(\mathbf{r}) -\varepsilon Z_s^2(\mathrm{r})$ with $s=P,AP$ stacking.
\begin{align}
V_s(\mathrm{r})&=\sum_{n=1}^3 
 \left[(A_1 e^{-Q d_0}\mathrm{cos}(\varphi_n(\mathbf{r})) + \nonumber\right.\\
&\qquad    \left. A_2 e^{-G d_0}  \mathrm{sin}(\varphi_n(\mathbf{r})  + \varphi_s)\right] \\
Z_s(\mathrm{r}) &= 
 \frac1{2 \epsilon} \sum_{n=1}^3 \left[Q A_1 e^{-Q d_0} \mathrm{cos}(\varphi_n(\mathbf{r}))+\right. \nonumber\\
&\qquad \left.     G A_2 e^{-G d_0} \mathrm{sin}(\varphi_n(\mathbf{r}) + \varphi_s)\right].
\end{align}
Here, $\varphi_n(\mathbf{r})=\mathbf{g}_n\cdot\mathbf{r}+  \mathbf{G}_n\cdot(u^{(t)}(\mathbf{r}) - u^{(b)}(\mathbf{r}))$ with 
$\mathbf{g}_n = \delta \mathbf{G}_n - \theta \mathbf{e}_z \times \mathbf{G}_n$ with $\mathbf{G}_n$ being the vectors of the first star of the reciprocal lattice $\pm \mathbf{G}_n (n=1,2,3)$ with $G=|\mathbf{G}_n|=\frac{4\pi}{\sqrt{3} a}$ (cf. Fig.  \ref{illustrationinterlayerX} b)), $Q=\sqrt{G^2+\rho^{-2}}$. Parameters are the lattice parameter $a$, $\varphi_{AP}=0$, and $\varphi_{P}=\pi/2$.
Parameters $\varepsilon$, $A_1$, $A_2$, \red{minimum of the potential $W_s$ in z-direction} $d_0$, and $\rho$ were obtained in \cite{PhysRevLett.124.206101} using a fitting procedure. $\theta$ is the local stacking angle, and $\delta$ is the parameter for the lattice mismatch of the two layers.
Used material parameters are given in table \ref{parametertable}.

For a calculation of the local in-place displacement $\mathbf{u}_{t/b}$, the integral in $\mathcal{E}$ is discretized on a regular grid along the unit cell basis vectors using finite differences. $\mathcal{E}$ is minimized using the TAO package from PETSC \cite{petsc-efficient,petsc-web-page,petsc-user-ref} using the BLMVM algorithm. Mathematica \cite{Mathematica} analytically calculated the required gradients on the grid for the optimization. We started the minimization process in the procedure from different random distributions with strain islands of different sizes and magnitudes.
The result seems stable to the naked eye against initial random strain field variations.
However, minor variations on a larger scale than the Moiré pattern unit cell remain in the supercell case after minimization. The case is similar to a not totally but almost relaxed structure - a typical experimental situation even for a nearly pristine prepared sample. These variations are only visible in optical spectra, resulting in inhomogeneous lineshapes, but allow us to incorporate disorder.
\red{The two monolayers can be stacked using parallel (P) or anti-parallel (AP) orientation of the unit cells.}
For the first example, we choose a sample with 0.9° rotation in P stacking of a MoSe$_2$-WSe$_2$ bilayer. Therefore, the interpretation in this paper is restricted to P stacking; the overall structure and physics of states in AP stacking can differ significantly.
Fig. \ref{strainandpots}(a) shows the bottom interlayer displacement of an $8\times 4$ supercell after relaxation. We see a strong reconstruction around the corners of a triangular shape visible in the interlayer strain plotted in Fig. \ref{strainandpots} (b). The triangular shape in the x and y component of the displacement (not shown) of $u^{(\cdot)}(\mathbf{r})$ as predicted in \cite{PhysRevLett.124.206101} is clearly visible also in the strain field Fig. \ref{strainandpots} (b)  and shows no immediately visible deviation from a $1\times 1$ Moiré cell. The area captured in optical experiments is typically 7-8 times larger than the area of the $8\times 4$ supercell. Therefore, later, an averaging of calculated spectra from multiple computational domains is necessary to capture typical experimental signals.

\section{(Interlayer-) Exciton states in 2D structures} \label{sec:interlayer_exc_states}
This paper will describe the dynamics of interlayer excitons in monolayer heterostructures under the influence of strain, including the influence of phonons and radiative decay.
As a first step to cover the quantum dynamics, we focus on the interlayer excitons between an electron in K valley in one layer and holes in K valley in another layer influenced by strain cf.  Fig. \ref{illustrationinterlayerX}. Intervalley scattering and including multiple exciton types (such as intralayer excitons) is subject to future studies.

The interlayer exciton wave function $\Psi(\mathbf{r}_e,\mathbf{r}_h)$ for two fixed bands obeys a two dimensional Schrödinger equation:
\begin{align}
E\Psi(\mathbf{r}_e,\mathbf{r}_h)=&\left[ -\frac{\hbar^2 \Delta_e}{2m_e}-\frac{\hbar^2 \Delta_h}{2m_h}+ V_{eh}(\mathbf{r}_e-\mathbf{r}_h)\right.\nonumber\\
&\left.\quad+V_h(\mathbf{r}_h)+V_e(\mathbf{r}_e)\right]\Psi(\mathbf{r}_e, \mathbf{r}_h),
\end{align}
with the energy $E$ of the exciton, the in-plane coordinates $\mathrm{r}_{e/h}$ of electron and hole, and effective masses $m_{e/h}$ of electron and holes forming the selected interlayer exciton.
$V_{eh}(\cdot)$ is the Coulomb interaction between electron and hole,
where a Rytova-Keldysh potential was used together with the interpolation formula from \cite{PhysRevB.84.085406} and the modifications for hetero-layers \cite{PhysRevB.102.195403}:
\begin{align}             
 V_{eh}(\mathbf{r})=\frac{e^2}{4 \pi \varepsilon_0 \varepsilon_r}   \frac{1}{r_{eff}}  \left[\mathrm{log}\left(\frac{r}{r+r_{eff}}\right)+(\gamma-\mathrm{log}(2))e^{-\frac{r}{r_{eff}}}\right].         
\end{align}
Here, $r_{eff} =r_*+r_*' +d$ and $r_*$ and $r_*'$ are the material parameters for the two layers, which are related to the in-plane electric susceptibilities \cite{PhysRevB.102.195403} and $\gamma$ is Euler's constant.

The single particle potentials $V_h(\mathbf{r}_h)$, $V_e(\mathbf{r}_e)$ result from different physical processes.
In quantum wells, interface and alloy fluctuations (in principle impurities) usually cause a random disorder potential \cite{ZIMMERMANN200389,PhysRevB.95.235307}. Here, for a 2D monolayer heterostructure, we focus on the influence of strain due to lattice reconstruction and some remains of an initial random strain field that is almost but not entirely relaxed. Analog to the quantum well case interface fluctuations, we initialize the strain field, assuming strain fluctuations of different island sizes and amplitudes.
Random fluctuation due to vacancies may be added in later studies.
Concerning strain, there are at least two contributions to the single particle potentials $V_{e/h}(\mathbf{r}_{e/h})=V_{e/h,s}^{inter}(\mathbf{r}_{e/h})+V_{e/h,s}^{intra}(\mathbf{r}_{e/h})+V_{e/h}^{conf}(\mathbf{r}_{e/h})$ together with a confinement potential useful in later studies.
The first contribution is caused by the relative displacement of the two layers of the heterostructure (interlayer contribution) and was derived in \cite{PhysRevB.97.035306,tran2019evidence} for the COM motion:
\begin{align}
V_{e/h}^{inter}(\mathbf{r}_{e/h})=  V_{exc} \sum_{n=1}^3 \mathrm{cos}(\varphi_n(\mathbf{r}_{e/h}) + \varphi_{exc,s})   
\end{align}
where we add to the ansatz from \cite{PhysRevB.97.035306,tran2019evidence}, that $\varphi_n(\mathrm{r})$ is also modified through $u^{(t/b)}(\mathbf{r})$. This contribution was missing in \cite{PhysRevB.97.035306,tran2019evidence} since \cite{PhysRevB.97.035306,tran2019evidence} did not include lattice reconstruction due to strain. 
Also, instead of the approach from \cite{PhysRevB.97.035306,tran2019evidence}, where the interlayer contribution is added on the exciton level,  we apply $V_{e/h}^{inter}(\mathbf{r}_{e/h})$ to the individual electron and holes (scaled by a factor $\frac{1}{2}$) also to include the smearing effect of the convolution with the relative wavefunction. The physical origin of this potential should be the individual carrier and not the exciton as a whole (although the distribution of $V_{exc}$ between hole and electron requires further studies).
Fig. \ref{strainandpots}(d)  shows for one $8\times 4$-super cell the interlayer strain contribution $V_{e/h}^{inter}(\mathbf{r}_{e/h})$ of the single particle contribution of an electron or hole. The contribution has a maximum of ten meV in the middle of the triangular area formed by reconstruction. In contrast, the contribution to the potential is minimal on the sides and corners of the triangles. The minima are relatively sharp (and form ridges between the triangle corners beside the minima at the corners), and the maxima are relatively flat and cover almost the whole triangular area. \red{The potential amplitude is similar to the result in Ref. \onlinecite{tran2019evidence}. However, the sharp potential minima in the triangle points are twice as deep due to reconstruction}.

Furthermore, the intralayer strain modifies the band gap offset of the electron or hole respectively inside the individual layers.
In \cite{Khatibi_2019}, the influence of intralayer strain on the band gap, effective masses of electrons, and holes in monolayers were calculated. The impact on effective masses was negligible compared to the effect on the band gaps. Therefore, for electrons and holes in the respective layer, we calculate a potential based on the intralayer-induced change of the local band gap energy as:
\begin{align}
    V_{e/h}^{intra}(\mathbf{r}_{e/h}) =\Delta E_{gap}^{e/h} \left(\frac{\partial u_x}{\partial x}+\frac{\partial u_y}{\partial y}\right)
\end{align}
assuming isotropic strain.
Note that electron and hole reside in different materials, so $\Delta E_{gap}^{e/h}$ provides only a rough estimate since \cite{Khatibi_2019} provides only a parameter for the band gap, i.e., the difference between electron and hole at K point. (Since both valence and conduction band shift to negative for tensile strain, this translates to negative $\Delta E_{gap}^{e} $ and positive $\Delta E_{gap}^{h}$).
For the $8\times 4$ supercell, we see in Fig. \ref{strainandpots} (c) also a triangular-shaped intra-layer strain potential $V_{e/h}^{intra}(\mathbf{r}_{e/h})$.
The potential is positive in the order of 10 meV in the middle of the triangular areas. In the edges and corners of the triangles, the potential reaches a negative minimum in the order of 10-20 meV.
However, the intralayer potential transition from minima to maxima is relatively smooth compared to the interlayer potential.
\red{Intralayer strain manifests in a piezoelectric polarization of the layers, which will also influence the carriers in the other layer \cite{PhysRevLett.124.206101,enaldiev2022self} leading to an additional interlayer contribution. However, to our knowledge, currently, no terms and parameters available in the literature allow a treatment independent from the intralayer term based on DFT. Therefore, we do not include the piezoelectric contribution and may underestimate the strength of the potential.}
Altogether, the contributions to the single particle potential can reach differences between minimum and maximum around 40-60 meV.
However, the areas with minimal potential remain slim and will have limited confined exciton states.
$V_{e/h}^{conf}(\mathbf{r}_{e/h})$ is a confinement potential that is zero inside the structure and has a finite value outside the structure. It will be used in future studies for non-infinitely extended structures.

\red{While we restrict the analysis to bright interlayer excitons, also interlayer couplings can be modulated by strain, which is, for example, discussed in \cite{PhysRevMaterials.4.094002,PhysRevLett.125.266404} using a description in dressed states that behave like particles in an effective magnetic potential.
Future studies may also include transfer from intralayer excitons to interlayer excitons via an interlayer coupling. As a result, we expect a strain-modulated interlayer coupling that results in a position-dependent transfer of carriers, especially affecting the transfer of more localized exciton states.}

Due to the strong Coulomb interaction in TMDCs, the exciton wave function is factorized into relative and center-of-mass \red{(COM)} movement $\Psi(\mathbf{r}_e,\mathbf{r}_h)=\psi(\mathbf{R})\phi(\mathbf{r})$ with the COM coordinate $\mathbf{R}=(m_e\mathbf{r}_e+m_h\mathbf{r}_h)/(m_e+m_h)$ and relative coordinate $\mathbf{r}=\mathbf{r}_h-\mathbf{r}_e$. 
Separating the Schrödinger equation into the relative and COM coordinates leads to the Wannier equation for the relative part of the wavefunction:
\begin{align}
E\phi(\mathbf{r})=\left[-\frac{\hbar^2 \Delta_r}{2m_r}+V_{eh}(\mathbf{r})\right]\phi(\mathbf{r}) 
\end{align}
with the relative mass $1/m_r=1/m_e+1/m_h$.
The Wannier equation is solved numerically using Petsc \cite{petsc-efficient,petsc-web-page,petsc-user-ref}, and Slepc \cite{Hernandez:2005:SSF} using finite differences for the partial differential equation. We restrict the following discussion to the 1s relative wave function and, thus, the lowest interlayer exciton states.
For the discussed example in this paper, we consider a K-K interlayer exciton with the electron residing in the top WSe$_2$ layer and the hole in the bottom MoSe$_2$ layer (cf. Fig. \ref{illustrationinterlayerX}).
The Schrödiger equations of the exciton's \red{COM} wave function $\Psi_{COM}(\mathbf{r})$ has a local potential $V_{com}(\mathbf{R})$:
\begin{align}
    \left(-\frac{\hbar^2\Delta_R}{2M} + V_{com}(\mathbf{R})\right) \psi(\mathbf{R}) = E \psi(\mathbf{R}).
\end{align}
where the local strain and confinement potential influence enters through $V_{e/h}(\mathbf{r}_{e/h})$.

This exciton COM wave function potential $V_{com}(\mathbf{R})$ is connected to the electron and hole potentials $V_{e/h}(\mathbf{r}_{e/h})$ as follows \cite{ZIMMERMANN200389,PhysRevB.95.235307}:
\begin{align}
V_{com}(\mathbf{R})=\int d^2 r |\phi_{1s}(\mathbf{r})|^2\left[ V_{e}\left(\mathbf{R}-\frac {m_h}{M}\mathbf{r}\right) +V_{h}\left(\mathbf{R}+\frac {m_e}{M}\mathbf{r}\right)\right].
\end{align}

\section{Computational domain and exciton states}
\label{sec:compdomianexcstates}
We use a periodic computational domain - supercell - consisting of several Moiré unit cells to calculate the COM wave function. We choose an integer multiple along the basis vectors of the Moiré unit cell to span the computational domain. 
For the boundary conditions of the strain, the strain can either match the non-reconstructed Moiré structure strain field at the boundary or have loose boundary conditions. The COM exciton potential would be set to a high positive finite value in parts with no material. 

To cover all the relevant dynamics inside the spectral range, we typically calculate up to 2500 COM excitons. The 2500 COM excitons wave functions were calculated within a feasible computation time of several days to a week. The covered spectral range varies depending on the size of a Moiré supercell. As a rule of thumb, the calculated exciton states should at least cover a spectral range one-third larger than the range of interest.
Furthermore, the real space COM wavefunctions allow a straightforward characterization of the exciton states contributing to peaks in spectroscopy.

\begin{figure}[tb]
    \centering
    \includegraphics[width=9cm]{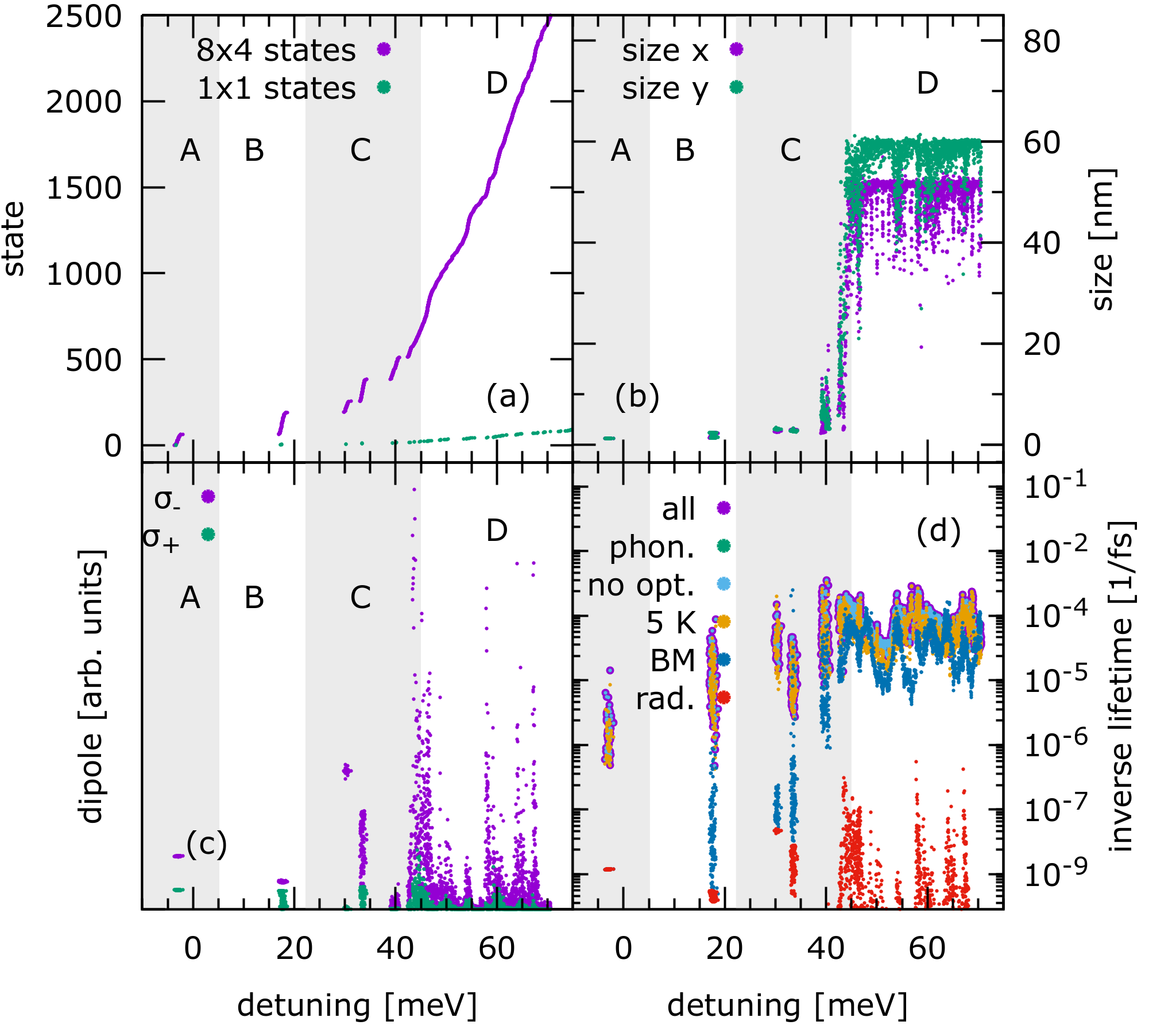}
    \caption{For the WSe$_2$/MoSe$_2$ heterolayer: (a) \red{state index} over energy for an 8x4 and 1x1 Moiré supercell, (b) size of the exciton states in x and y direction, (c) dipole moment component for $\sigma_+$ and $\sigma_-$heterostructure energy, (d) exciton lifetime at 20 K in polaron transformation including all scattering contributions (all), including only exciton-phonon scattering (phon.), excluding the optical phonon contribution (no opt.), at 5 K, without polaron transformation: standard Born Markov (BM)  and only including the radiative contribution. The states are sorted into energy ranges with common properties. This includes the first localized states in area (A), the second localized states in area (B), the higher energy localized states in area (C), and the transition at the mobility edge to the delocalized states in area (D).}
    \label{plots_A}
\end{figure}
\begin{figure}[tb]
    \centering
    \includegraphics[width=8.6cm]{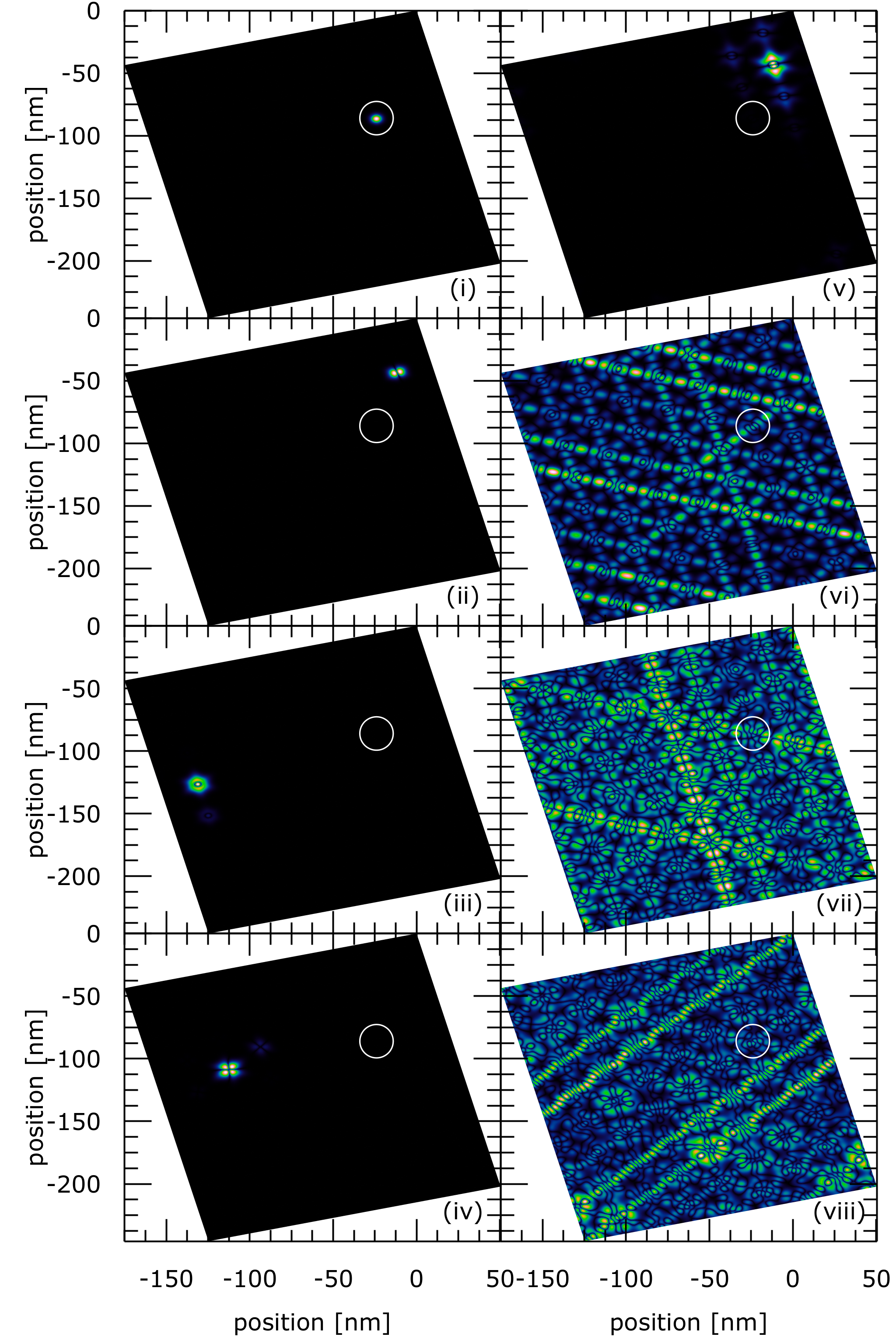}
    \caption{For the WSe$_2$/MoSe$_2$ heterolayer for an 8x4 and 1x1 Moiré supercell, (i) state 1 at \red{-3.57 meV}, (ii) state 98 at 17.38 meV, (iii) state 210 at 30.12 meV, (iv) state 280 at 33.22 meV, (v) state 410 at \red{39.35 meV}, (vi) state 1000 at 49.11 meV, (vii) state 2000 at 64.05 meV and (viii) state 2320 at 67.88 meV (energy relative to the bound interlayer exciton energy). \red{The white circle marks the position of state $1$ for visual orientation.}}
    \label{plots_B}
\end{figure}
{\it Illustrative example of the excitons in a heterolayer:} We use the example of a MoSe$_2$/WSe$_2$ heterolayer with P stacking and a twist angle of 0.7° to discuss the exciton structure. 
Note that the interlayer component of the exciton potential $V_{exc}$ is ten times stronger in P stacking than in AP stacking, so a more quantum dot-like behavior will be observed in the selected example with P stacking. In AP stacking, the exciton states will be more delocalized. However, a discussion of this case is beyond the scope of this article.
If not mentioned otherwise, we use an $8\times4$ supercell and one random realization for this example. 
In Fig. \ref{plots_A}a) the exciton state energy over the index of the exciton state is plotted (lower energy to higher energy), we see different steps in the plot separating the exciton states of different characters, which form different peaks in the spectrum. The overall qualitative structure is independent of the random realization.
These peaks may be interpreted as different mini bands if the simulation does not contain disorder. The peaks can also be related to levels of localized (an-)harmonic oscillators \cite{PhysRevB.97.035306} for the two lower peaks. The third peak contains contributions from several higher localized oscillator levels and the transition to delocalized states at the mobility edge.
The energetically lowest set of exciton states (area A and B in Fig. \ref{plots_A}) are completely localized anharmonic oscillator states, as can be seen in the absolute square of some selected COM wave functions plotted in Fig. \ref{plots_B} (i) and (ii). The states in area A correspond to the lowest s-type state, and the states in area B to the higher p-type localized states. In area C, below the mobility edge, various higher localized anharmonic oscillator states are present (see exemplary states  Fig. \ref{plots_B} (iii) to (v)). The delocalized states start at the mobility edge starting at area D. As seen in Fig. \ref{plots_B} (vi), the lower delocalized states are still limited to the ridges of the structure, only the higher delocalized states (see Fig. \ref{plots_B} (vii) and (viii) for example) are freely distributed over the structure. Also, the long-range modulation of the strain-induced potential is visible in these states, but of course, it also causes the inhomogeneous distribution of the lower states. 
The patterns found in the delocalized states are diverse and include states concentrated at the triangle corners, others focused at one edge of the triangle, or others concentrated at the other edge. Also, various superpositions of these states, all spatially modulated by the remaining strain disorder, are present.
The mobility edge can also be seen in Fig. \ref{plots_A}(b), where the size of the COM wavefunctions is plotted over energy.
However, due to a long-range modulation of the strain, the mobility edge is blurred, and in part of area D, predominantly localized states exist, as can be seen in Fig. \ref{plots_A} (b).
 The nature of the localized states changes compared to earlier calculations without reconstruction and remaining disorder, in which for interlayer excitons \cite{PhysRevB.97.035306}, the wave function showed the same symmetries but were rather sharp lines and spread over all Moiré unit cells. Here, an inhomogeneous distribution is created.

\section{Exciton scattering}
\label{sec:excitonscattering}
\subsection{Exciton-phonon scattering}
Exciton-phonon scattering is the major component determining the exciton dynamics in our setup.
It is described by an electron-phonon Hamiltonian (in second quantization):
\begin{align}
    H_{e-ph}=\sum_{q\lambda k \alpha G} g^{\lambda}_{q+G\alpha}a^\dagger_{\lambda k+q}a_{\lambda k}(b_{q\alpha}+b^\dagger_{-q\alpha}), \label{electron-phonon-coupling}
\end{align}
with the electron creation $a^\dagger_{\lambda k}$ and annihilation operator $a_{\lambda k}$ \red{with quasi momentum $k$} for valence $\lambda=v$ and conduction $\lambda=c$ band in electron picture and the phonon creation $b^\dagger_{q\alpha}$ and annihilation operators $b_{q\alpha}$ for phonons of mode $\alpha$ and quasi momentum $q$.
\red{Please note, the expression using quasi momenta $q$ and $k$ in Eq. \eqref{electron-phonon-coupling} appears before transforming the exciton wave functions into position space. The formulation in position space does not require a restriction to mini bands, nor is it allowed as disorder order breaks the discrete translation invariance on the Moiré level. }

We focus on acoustical phonons with deformation coupling \cite{PhysRevB.90.045422}:
\begin{align}
    g^{\lambda}_{q\alpha,ac}=\sqrt{\frac1{\hbar2\rho v_A q A}} D_{1\alpha }^{\lambda}|q|. \label{electron_phonon}
\end{align}
with the density $\rho_{e/h}$ (see table \ref{parametertable}) of the top or bottom layer, the speed of sound $v_a$ the deformation potential of electron and holes $D_e^{1\alpha}$, $D_h^{1\alpha}$ of their respective materials and optical phonons also with deformation coupling \cite{PhysRevB.90.045422}:
\begin{align}
    g^{\lambda}_{q\alpha,opt}=\sqrt{\frac1{\hbar2\rho \omega_{q\alpha} A}} D_{0\alpha}^{\lambda}, \label{electron_phonon_opt}
\end{align}
again with the deformation potential of the respective layer.

A derivation similar to \cite{ZIMMERMANN200389} leads to a Hamiltonian for the electron-phonon coupling in a single exciton basis ($|\alpha\rangle$):
\begin{align}
    H_{ex-ph}=\sum_{\alpha\beta q \tilde{\alpha} G } t_{\alpha\beta}^{q G \tilde{\alpha}} (b_{q\tilde{\alpha}}+b_{-q\tilde{\alpha}}^\dagger) |\alpha\rangle\langle\beta|
\end{align}
with matrix elements
\begin{align}
    t^{qG\tilde{\alpha}}_{\alpha\beta}=&\hbar g^{e}_{q+G\tilde{\alpha}} \int\mathrm{d}^2 r|\phi_{1s}(\mathbf{r})|^2 e^{-i\mathbf{r}\cdot\frac{ (\mathbf{q}+\mathbf{G})}{\beta_e}}\nonumber\\ 
    &\quad \int\mathrm{d}^2 R\psi_{\alpha}^*(\mathbf{R})e^{-i(\mathbf{q}+\mathbf{G})\cdot\mathbf{R}}\psi_{\beta}(\mathbf{R}) \nonumber\\
    &-\hbar g^{h}_{q+G\tilde{\alpha}} \int\mathrm{d}^2 r|\phi_{1s}(\mathbf{r})|^2 e^{-i\mathbf{r}\cdot \frac{(\mathbf{q}+\mathbf{G})}{\beta_h}}\nonumber\\ 
    &\quad \int\mathrm{d}^2 R\psi_{\alpha}^*(\mathbf{R})e^{-i(\mathbf{q}+\mathbf{G})\cdot\mathbf{R}}\psi_{\beta}(\mathbf{R}) \label{electron_phonon_excitons}
\end{align}
with $\beta_{e}=M/m_h$ and $\beta_h=M/m_e$.
Please note, if electron and holes are not residing in the same monolayer, a different phonon mode branch is acting on the electron and hole so that only one of $g^{e}_{q+G\tilde{\alpha}}$ or $g^{h}_{q+G\tilde{\alpha}}$ is non-zero for a given phonon mode.
The phonon branch index is included as a multi-index inside $q$.
\red{We still included all $q$ indices. Effects like zone folding for $q$ that appear for strict periodic Moiré unit cells would enter implicitly through the COM exciton wavefunction $\psi_{\alpha}^*(\mathbf{R})$ and does appear explicitly in the matrix elements for the formulation in position space. However, as disorder is present effects of zone folding only occur approximately.}

We can rewrite Eq. \eqref{electron_phonon_excitons} using form factors $\chi(\mathbf{q})=\int d^2 r |\phi_{1s}(\mathbf{r})|^2 e^{-i \mathbf{q}\cdot\mathbf{r}}$ and Fourier transforms of the overlap of the COM states
$O_{\alpha\beta}(\mathbf{q})=\int\mathrm{d}^2 R\psi_{\alpha}^*(\mathbf{R})e^{-i(\mathbf{q}+\mathbf{G})\cdot\mathbf{R}}\psi_{\beta}(\mathbf{R}) $:

\begin{align}
    t^{qG\tilde{\alpha}}_{\alpha\beta}=&\hbar g^{e}_{q+G\tilde{\alpha}} \chi\left(\frac{\mathbf{q}+\mathbf{G}}{\beta_e}\right) O_{\alpha\beta}(\mathbf{q}+\mathbf{G}) \nonumber\\
    &-\hbar g^{h}_{q+G\tilde{\alpha}} \chi\left(\frac{\mathbf{q}+\mathbf{G}}{\beta_h}\right) O_{\alpha\beta}(\mathbf{q}+\mathbf{G})  \label{electron_phonon_excitons_rewritten}
\end{align}
\subsubsection{Fermi's golden rule}
Applying Fermi's golden rule to the exciton phonon interaction yields:
\begin{align}
    \gamma_{\beta\leftarrow\alpha}=&\frac{2\pi}{\hbar} \sum_{q\tilde{\alpha}}|t^{q\tilde{\alpha}}_{\alpha\beta}|^2(n(\hbar \omega_q)\delta(\varepsilon_{\beta}-\varepsilon_{\alpha}-\hbar\omega_q) \nonumber\\
    &\qquad +(1+n(\hbar\omega_q))\delta(\varepsilon_{\beta}-\varepsilon_{\alpha}+\hbar\omega_q)), \label{fermisgoldenruleeq}
\end{align}
with the Bose-Einstein distribution $n(\cdot)$.
Since electron and hole would reside in different monolayers, the rates scale roughly $|D^e_1|^2+|D^h_1|^2$ (omitting some electron/hole-dependent prefactors for readability). In contrast, for an intralayer process within the same monolayer, the rate would scale as $|D^e_1-D^h_1|^2$, again without some electron and hole-dependent prefactors, in front of the deformation potential.
The rate has the same form for all contributing phonon modes, including the acoustic phonons LA and TA and the usual optical phonons LO, TA, and A1.
For the numerical implementation, we use $\hbar \omega_{q}=\hbar v_{a\alpha}|q|$ with the speed of sound $v_a$ for LA and TA phonons in the respective layer of the phonon mode, when applying the $\delta$-distibution.

A constant energy value often approximates the optical phonon's energy in Einstein's approximation. We include here a linear correction: $\hbar \omega_{ph,\alpha,\mathbf{q}}=\omega_{ph,\alpha,\mathbf{q}}^{(0)}+ \omega_{ph,\alpha,\mathbf{q}}^{(1)} q$, where $\omega_{ph,\alpha,\mathbf{q}}^{(1)}$ is determined by the difference in phonon energy at the $\Gamma$- and $Q$- point \red{(cf. Fig. \ref{illustrationinterlayerX}b))}. Thus, the phonon energies for different $\mathbf{q}$ form a narrow continuum.

As we will see later, the linewidth of localized states in absorption spectra 
is predominantly caused by the multi-phonon processes (mostly acoustic with optical satellite peaks) included by polaron transformation, which we introduce in the following subsection. The relaxation processes are relatively slow and give only a negligible contribution. On the other hand, optical phonons are crucial for the long-time dynamics between localized states at different positions because of the large energy spacing.

\subsubsection{Polaron transformation and modified Fermi's golden rule}
So far, Fermi's golden rule has only included single-phonon scattering processes.
Especially for energetically deep localized states, as we find here, multi-phonon processes might be necessary for a correct description of the system. Otherwise, we may see in the simulation the phonon-bottleneck effect (e.g., see discussion for quantum dots \cite{PhysRevB.59.5069}), i.e., that relaxation to localized states is effectively prohibited due to a mismatch between optical phonon energy and energy difference to the next states. 
In real systems, the phonon bottleneck effect is often softened or eliminated by scattering processes involving multiple phonons  (e.g., a mixture of optical and acoustic).
For localized  and energetically well-separated exciton states, especially the exciton-phonon coupling
diagonal in the exciton states (cf. in Eq. (\ref{electron_phonon_excitons}) $\alpha=\beta$) is essential for the exciton lineshape and also for modifying/softening energy conservation in the $\delta$-distribution in Eq. (\ref{fermisgoldenruleeq}).
Thus, we will carry out a polaron transformation for diagonal coupling on the Hamiltonian to include multi-phonon processes.

A polaron transformation \cite{mahan2000many,PhysRevB.57.347,PhysRevB.65.235311, PhysRevB.85.115309,PhysRevB.93.155423} is a canonical transformation, that transforms any operator $A$ as 
\begin{align}
    A'=e^{\sum_{\alpha} s_\alpha} A e^{-\sum_{\alpha} s_\alpha}
\end{align}
where we include only diagonal couplings in the generating operators
\begin{align}
s_\alpha =|\alpha\rangle\langle\alpha| \sum_{qG\tilde{\alpha}}\frac{t_{\alpha\alpha}^{qG\tilde{\alpha}}}{\hbar \omega_{q\tilde{\alpha}}} (b^\dagger_{-q\tilde{\alpha}}-b_{q\tilde{\alpha}}).
\end{align}
The transformation of the phonon and exciton Hamiltonian shift the exciton energies by the polaron shift $E'_\alpha=E_\alpha - \sum_{qG\tilde{\alpha}}\frac{|t_{\alpha\alpha}^{qG\tilde{\alpha}}|^2}{\hbar \omega_{q\tilde{\alpha}}}$.
The transformation removes the diagonal exciton-phonon coupling Hamiltonian and results in a transformed off-diagonal exciton-phonon Hamiltonian:
\begin{align}
    H'_{ex-ph}=\sum_{\alpha\neq\beta q \tilde{\alpha} G } t_{\alpha\beta}^{q G \tilde{\alpha}} B_+^\alpha (b_{q\tilde{\alpha}}+b_{-q\tilde{\alpha}^\dagger}) B_-^\beta |\alpha\rangle \langle\beta|,
\end{align}
where the operators $B^\alpha_{\pm}$ change the phonon equilibrium position from one exciton state $\alpha$ to $\beta$:
\begin{align}
B^\alpha_{\pm} = \mathrm{exp}\left(\pm \sum_{qG\tilde{\alpha}} \frac{t_{\alpha\alpha}^{qG\tilde{\alpha}}}{\hbar \omega_{q\tilde{\alpha}}} (b^\dagger_{-q\tilde{\alpha}}-b_{q\tilde{\alpha}}) \right).
\end{align}
$H'_{ex-ph}$ does not include only the dynamic contribution of the exciton-phonon interaction but also a static off-diagonal tunnel-like contribution.
For the calculation of transfer rates similar to Fermi's golden rule, we separate the static contribution from the dynamic contribution: $H''_{ex-ph}+H^{static}_{ex-ph}=H'_{ex-ph}$:
\begin{align}
H''_{ex-ph}= &\sum_{\alpha\neq\beta q \tilde{\alpha} G } t_{\alpha\beta}^{q G \tilde{\alpha}} |\alpha\rangle \langle\beta|  B_+^\alpha \zeta_{\alpha\beta}^{qG\tilde{\alpha}} B_-^\beta  , \\
H^{static}_{ex-ph}=& \sum_{\alpha\neq\beta q \tilde{\alpha} G } t_{\alpha\beta}^{q G \tilde{\alpha}} |\alpha\rangle \langle\beta| B_+^\alpha ( t_{\alpha\alpha}^{-q G \tilde{\alpha}}+t_{\beta\beta}^{-q G \tilde{\alpha}}) B_-^\beta \nonumber
\end{align}
with the shifted phonon operators $\zeta_{\alpha\beta}^{qG\tilde{\alpha}} = b_{q\tilde{\alpha}}+b_{-q\tilde{\alpha}}^\dagger-t_{\alpha\alpha}^{-q G \tilde{\alpha}}-t_{\beta\beta}^{-q G \tilde{\alpha}}$, which accounts for the shift of the equilibrium position.
The static contribution accounts for the shift of the potential surface caused by the transition from one exciton state to another. We assume its effects were included in the underlying band structure or negligible in line with previous treatments using a polaron transformation \cite{PhysRevB.85.115309}.

To obtain the analog of Fermi's golden rule for the polaron-transformed Hamiltonian, the temporal correlation function of $H''_{ex-ph}$ is evaluated:
\begin{align}
&\mathrm{tr}_B(\langle \alpha|H''_{ex-ph}(t)|\beta\rangle \langle \beta| H''_{ex-ph}(0) |\alpha\rangle \rho_B)=\nonumber\\&\qquad\qquad\qquad \mathrm{exp}({i (\varepsilon_\beta -\varepsilon_\alpha)t/\hbar + G_{\alpha\beta}(t)})H_{\alpha\beta} (t),  \\
&G_{\alpha\beta}(t)=\sum_{q \tilde{\alpha} G }\frac{|t_{\alpha\alpha}^{q G \tilde{\alpha}}-t_{\beta\beta}^{q G \tilde{\alpha}}|^2}{\hbar^2 \omega_{q\tilde\alpha}^2}(n_{q \tilde{\alpha}}(e^{i \omega_{q\tilde\alpha} t}-1) \nonumber\\&\qquad\qquad\qquad+(n_{q \tilde{\alpha}}+1)(e^{-i \omega_{q\tilde\alpha} t}-1)),\\
&H_{\alpha\beta}(t)=\sum_{ q \tilde{\alpha} G } |t_{\alpha\beta}^{q G \tilde{\alpha}}|^2 (n_{q \tilde{\alpha}}e^{i \omega_{q\tilde\alpha} t} +(n_{q \tilde{\alpha}}+1)e^{-i \omega_{q\tilde\alpha} t})\nonumber\\
&\quad+(\sum_{ q \tilde{\alpha} G } t_{\alpha\beta}^{q G \tilde{\alpha}} t_{\alpha\alpha\beta\beta}^{-q G \tilde{\alpha}} (n_{q \tilde{\alpha}}e^{i \omega_{q\tilde\alpha} t} -(n_{q \tilde{\alpha}}+1)e^{-i \omega_{q\tilde\alpha} t}))\nonumber\\
&\quad\quad \times (\sum_{ q \tilde{\alpha} G } t_{\beta\alpha}^{q G \tilde{\alpha}}t_{\alpha\alpha\beta\beta}^{-q G \tilde{\alpha}} (n_{q \tilde{\alpha}}e^{i \omega_{q\tilde\alpha} t} -(n_{q \tilde{\alpha}}+1)e^{-i \omega_{q\tilde\alpha} t}))
\end{align}
with $t_{\alpha\alpha\beta\beta}^{-q G \tilde{\alpha}} = t_{\alpha\alpha}^{-q G \tilde{\alpha}}-t_{\beta\beta}^{q G \tilde{\alpha}}$ and
where $n_{q \tilde{\alpha}}$ is the Bose distribution with energy $\hbar \omega_{q\tilde\alpha}$.
Here, we assumed an initial harmonic bath and used cumulant expansion in the second order of the phonon couplings for the calculation. The second-order cumulant expansion is exact, as Wick's theorem holds. $G_{\alpha\beta}(t)$ accounts for the spectral phononic overlap of the initial and final states in the transition from $\alpha$ to $\beta$. If the final and initial exciton states have the same interaction with the phonon environment, then $G_{\alpha\beta}(t)$ will vanish as the phononic lineshape has the same form. The first term at $B_{\alpha\beta}(t)$ describes a typical transition from exciton state $\alpha$ to state $\beta$, including the non-diagonal exciton phonon interaction in second order, where the phonon emitted or absorbed at the first interaction is also involved in the second interaction. The second term is different as a different phonon is involved in each off-diagonal interaction as the process involves additional diagonal interaction. As in the exponential, the second term will only contribute if the interaction of the initial and final state differs.
Using a second-order quantum master equation in the off-diagonal exciton phonon coupling, we can extract a polaron scattering rate:
\begin{align}
&\gamma_{\beta\leftarrow \alpha}=\int_0^\infty \frac{e^{-\gamma t}}{\hbar^2}\mathrm{tr}_B((H''_{ex-ph}(t))_{\alpha\beta}( H''_{ex-ph}(0) )_{\beta\alpha}\rho_B) \mathrm{d}t , \label{exciton_phonon_rate_polaron}
\end{align}
with $A_{\alpha\beta}=\langle\alpha|A|\beta\rangle$,
where we added a phenomenological $\gamma=0.0001\mathrm{eV}$ to achieve numerical convergence (and is required to approximate the $\delta$ distribution)in case the phononic environment does not substantially differ between states $\alpha$ and $\beta$).

\begin{figure}[bth]
    \centering
    \includegraphics[width=8.6cm]{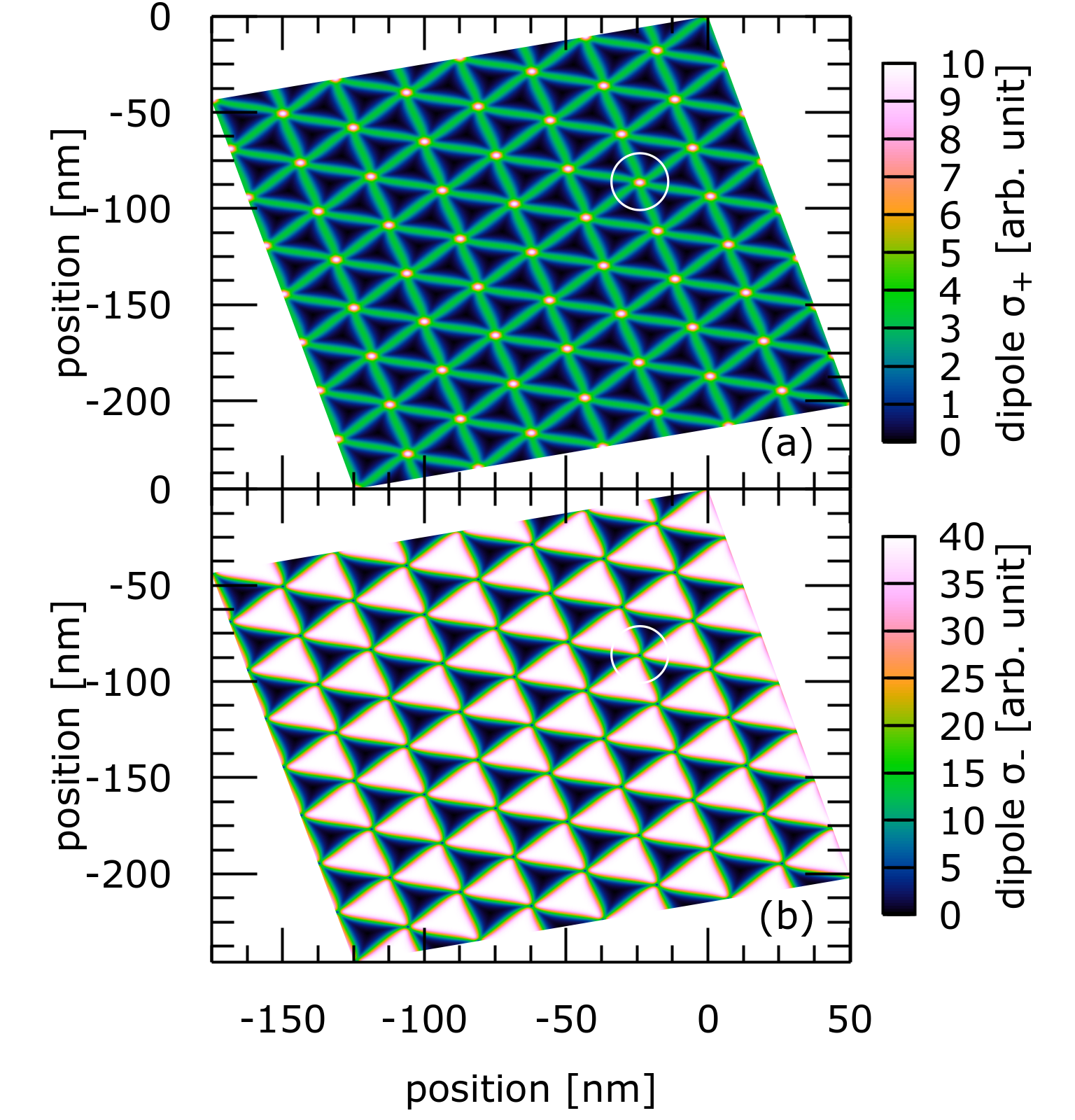}
    \caption{For the WSe$_2$/MoSe$_2$ heterolayer example in an 8x4 supercell, the absolute value of dipole element for the (a) $\sigma_+$ and (b) $\sigma_-$. Arbitrary units but scale to the ratio of microscopical elements from \cite{PhysRevB.97.035306}. \red{The white circle marks the position of state $1$ for visual orientation.}}
    \label{dipolemomentplot}
\end{figure}
\subsection{Radiative decay}
Besides the exciton-phonon interaction, exciton light recombination is another possible scattering event.
The dipoles of the interlayer excitons are tiny compared to the single monolayer excitons and act on a longer time scale \cite{nagler2017interlayer}, so their influence is almost negligible to exciton dynamics on the time scales relevant for exciton-phonon dynamics.
In general, the starting point is the electron-photon interaction.
To include the influence of reconstruction, we build upon \cite{PhysRevB.97.035306}.
Analog to \cite{PhysRevB.97.035306} (Eq. (8) ibid), we introduce a parametrization of the current:
\begin{align}
\mathbf{D}(\mathbf{r})&= 
 \frac{D_+}{3} (1 + e^{+i \varphi_3(\mathbf{r})} + e^{-i \varphi_2(\mathbf{r})} ) \mathbf{e}_+\nonumber \\
&+
 \frac{D_-}{3} (1+ 
    e^{i \varphi_3(\mathbf{r}) - i \mathbf{G}_3\cdot r_{0,1}} + 
    e^{-i \varphi_2(\mathbf{r}) + i \mathbf{G}_2 \cdot r_{0,1} })\mathbf{e}_-
\end{align}
with $r_{0,n} :=   -(4 \pi) n/(3 G^2) \mathbf{G}_0$, where we replaced the global lattice displacement dependence from \cite{PhysRevB.97.035306} with local phase variation due to reconstruction.
Even in \cite{PhysRevB.97.035306} $\mathbf{D}(\mathbf{r})$ was defined as related to the current of the whole exciton. It is directly connected to the current matrix element $\mathbf{d}_{vc}(\mathbf{r})$ of a unit cell with: 
\begin{align}
\mathbf{d}_{vc}(\mathbf{r})=\int_{\Omega(\mathbf{r})} d^3r' u^*_{k\approx 0,v}(\mathbf{r}') \frac{e}{m} \mathbf{p} u_{k\approx 0, c} (\mathbf{r}')
\end{align}
with $\Omega(\mathbf{r})$ the volume of the unit cell at position $\mathbf{r}$ through $\mathbf{D}(\mathbf{r})=\frac{\mathbf{d}_{vc}(\mathbf{r})}{\Omega(\mathbf{r})}$.
Plots of $\sigma_-$ and $\sigma_+$ components of the $8\times 4$ supercell in Fig. \ref{dipolemomentplot}, respectively, again correlate to the triangular-shaped pattern in the Moiré cell. The $\sigma_-$ component shows a flat triangular shape, while the $\sigma_+$ component has the most prominent contribution along the edges of the triangle. The $\sigma_-$-component is much larger than the  $\sigma_+$-component, which can be traced back to the DFT parameters from \cite{PhysRevB.97.035306}. So, the valley selectivity is limited in this type of heterostructure, especially the localized states that reduce the selectivity. Of course, if the K valley is exchanged with the K' valley, the selection rule flips so that both signals are detected simultaneously for spectroscopic signals.

The oscillator strength of the calculated exciton states (cf. Fig. \ref{plots_A}(c)) behaves differently to the quantum well case, where the few localized lower energy states carry much higher individual oscillator strength than the higher energy delocalized states above the mobility edge. Here, we see a repetitive pattern in the delocalized states, where some delocalized states have a stronger dipole strength than the lower localized states, but some have a lower dipole strength. This can be traced back to the dipole moments in Fig. \ref{dipolemomentplot}. Specifically, the larger dipole moment for the $\sigma_-$ contribution has a minimum at the corners of the triangles, where the localized states reside. Furthermore, the $\sigma_+$ component is maximal at the corners, explaining the reduced valley selectivity of the localized states compared to the delocalized states.

Since the potential is relatively deep (40-60 meV for the P stacking case instead of a few meV typical for quantum wells), the localized state covers most of the visible optical features. The delocalized states in Area D cover only a minor portion of the visible spectrum. The localized states supply most other parts.

We can use the current parameterization to define the electron-photon Hamiltonian in rotating wave approximation:
\begin{align}
H_{el-ph} = -\sum_{k\sigma k_1 k_2} \sqrt{\frac{\hbar}{2\varepsilon_0 c \sqrt{\varepsilon_B} k V}} A^{k\sigma}_{k_1 k_2} b_{k\sigma}^\dagger  a^\dagger_{v \mathbf{k}_1}  a_{c \mathbf{k}_2} +h.a.
\end{align}
for a transition between valence band $\lambda=v$ and conduction band $\lambda=c$
with the coupling element:
\begin{align}
A_{k_1k_2}^{k\sigma}=\int \mathrm{d}^2R~ e^{i\cdot (\mathbf{k}_1-\mathbf{k}_2+\mathbf{k})\cdot R} \mathbf{e}_{\sigma}\cdot\mathbf{D}(\mathbf{R}).
\end{align}
A transformation to the exciton basis yields:
\begin{align}
&H_{ex-ph}=\sum_{k\sigma\alpha} \varphi_{1s}(0) \sqrt{\frac{\hbar}{2\varepsilon_0 c \sqrt{\varepsilon_B} k V}}  b^\dagger_{k\sigma}
\mathbf{e}_{\mathbf{k}\sigma}\cdot \nonumber\\ 
&\quad \int \mathrm{d}^2 r \mathbf{D}(\mathbf{r}) \psi_\alpha(\mathbf{r}) e^{i\mathbf{k}\cdot\mathbf{r}} |g\rangle\langle \alpha| + h.a.
\end{align}
Again analog to \cite{ZIMMERMANN200389}, we obtain an exciton-photon decay rate:
\begin{align}
r_\alpha =\frac{E_\alpha \sqrt{\varepsilon_B}}{2\varepsilon_0 c^3 \hbar^2} |\phi_{1s}(0)|^2 \left| \int \mathrm{d}^2 r \mathbf{e}_{k\sigma}^* \cdot \mathbf{D}(\mathbf{r}) \Psi_\alpha(\mathbf{r}) 
\right|^2, \label{eq:exciton_photon_decay}
\end{align}
assuming that the states are much smaller than the photon wavelength.

\begin{figure}[tb]
    \centering
    \includegraphics[width=8.9cm]{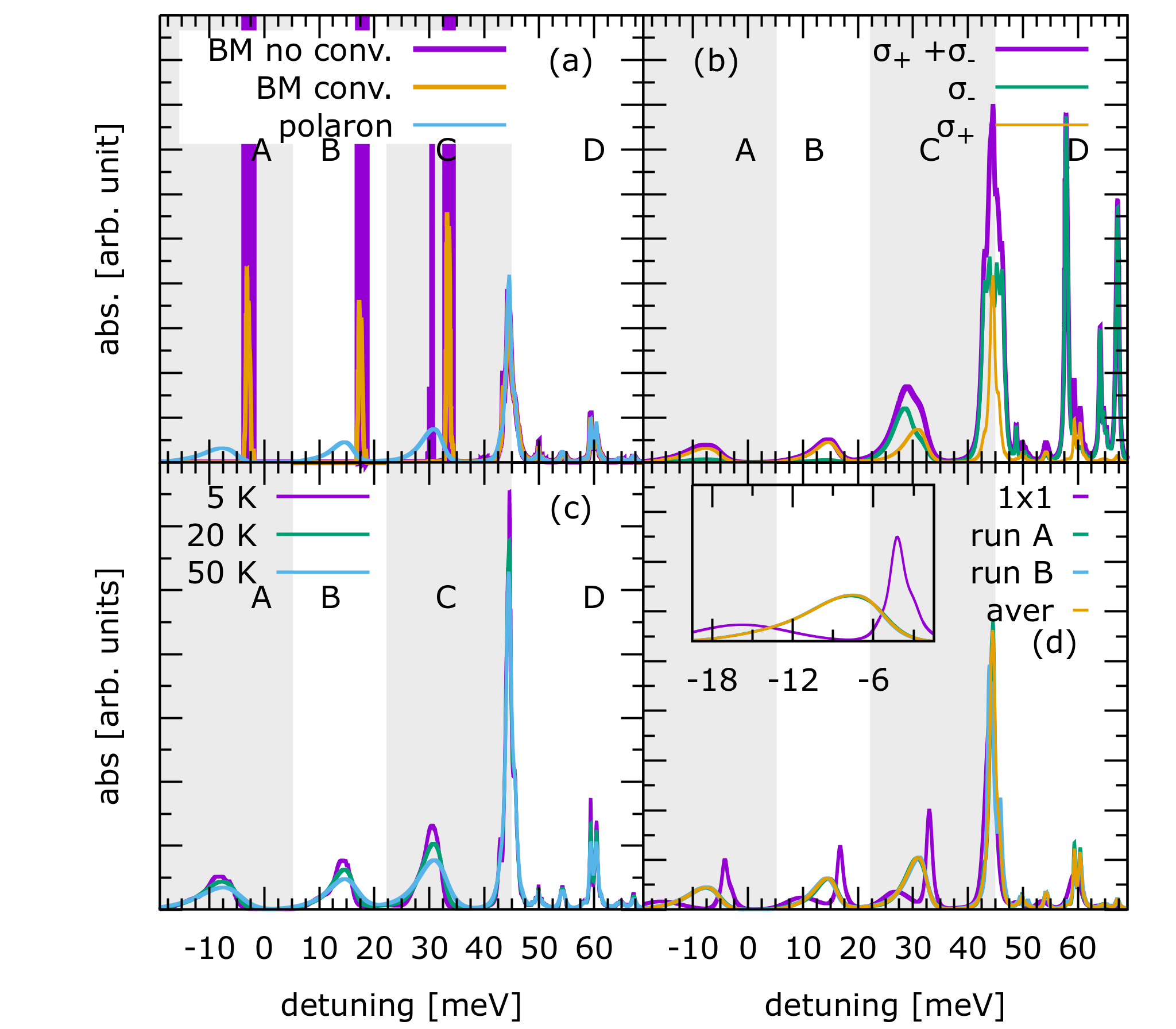}
    \caption{The WSe$_2$/MoSe$_2$ heterolayer linear spectra: for a single realization of an 8x4 supercell (a) spectrum at $20 K$ for $\sigma_+$ polarization in Born-Markov approximation with and without a convolution with Gaussian accounting for a typical spectral resolution of $0.15 \mathrm{meV}$ and using a polaron transformation ((b) through (d) use polaron transformation), (b) at $20K$ for $\sigma_-$ and $\sigma_+$ polarization and sum effectively collecting K and K' contributions, (c) for $\sigma_+$ polarization at different temperatures, (d) for $\sigma_+$ polarization comparison to a 1x1 supercell (scaled), two different random supercells (run A and run B) and averaging over 20 realizations. }
    \label{plots_C}
\end{figure}
\section{Quantum dynamics and numerical results}
\label{sec:eqmonumres}
The quantum dynamics accessed in optical spectroscopy are typically expressed in correlation functions \cite{mukamel1995principles}. For example, in linear spectroscopy, where at $t_0=0$ the exciton $\alpha$ is created from the ground state using $\sigma_\alpha^+=|\alpha\rangle\langle g|$ from the electron-light and interaction and then measured (destroyed through the microscopic polarization operator $\alpha^-=|g\rangle \langle\alpha'|$ for exciton $\alpha'$ at time $t$), the correlation functions $\mathrm{tr}(\sigma_\alpha^-(t) \sigma_{\alpha}^+(0)\rho)$   and $\mathrm{tr}(\sigma_\alpha^+(t) \sigma_{\alpha}^-(0)\rho)$ in Heisenberg picture contain all required information about linear absorption or luminescence (respectively) spectroscopy in secular approximation \cite{mukamel1995principles}.
If we use scattering rates without a polaron transformation, simple equations of motion for the density matrix elements are sufficient \cite{PhysRevB.95.235307}.
In this case, the correlation function can be evaluated using the  ground-to-one-exciton coherence $\rho_{\alpha g}$ dynamics, calculated using the equation of motion:
\begin{align}
\partial_t \rho_{\alpha g}=-i(E_\alpha-E_g-i\gamma_\alpha)\rho_{\alpha g},
\end{align}
where the single exciton decay rate $\gamma_\alpha$ is calculated using the rates using $\gamma_\alpha=\frac12(r_{\alpha}+\sum_{\alpha\neq \alpha'} \gamma_{\alpha\rightarrow \alpha'})$ the inverse lifetime of the exciton states.
In this approximation, the Green functions of the ground to excited state coherences are directly connected to the absorption and photoluminescence spectra.
As we will see, this description is insufficient to explain the lineshapes, especially of the localized states, so we must use the formulation in polaron transformation.
After polaron transformation $\sigma_\alpha^+$ and $\sigma_\alpha^-$ in the exciton-light interaction is replaced with $B^\alpha_+\sigma_\alpha^+$ and $\sigma_\alpha^-B^\alpha_-$, so that the correlations functions  $\mathrm{tr}(\sigma_\alpha^-(t)B^\alpha_-(t) B^\alpha_+(t)\sigma_{\alpha}^+(0)\rho)$ and $\mathrm{tr}(B^\alpha_+(t)\sigma_\alpha^+(t) \sigma_{\alpha}^-(0)B^\alpha_-(0)\rho)$ determine linear absorption and luminescence respectively.
The appearing $B^\alpha_\pm$ describes the nuclear reorganization from the diagonal exciton-phonon interaction after optical excitation, especially for localized lower energy states. We will see that it is the dominating contribution to the exciton lineshape.
In the polaron formulation, the remaining time propagation between $0$ and $t$ contains the homogenous exciton and phonon part $H_0$ of the Hamiltonian and non-diagonal exciton-phonon interaction $H''_{ex-ph}$.
Similar to Ref. \onlinecite{mahan2000many}, we assume that the non-diagonal exciton-phonon interaction is much weaker than the diagonal part (which should be valid for most of the localized states and is less critical for the delocalized states) and acts on longer time scales, where the nuclear reorganization is already finished.
Therefore, as an approximation, the time propagator in the correlation function is expanded in a second-order cumulant expansion \cite{renger2002relation} concerning non-diagonal exciton-phonon interaction $H''_{ex-ph}$, in the spirit of Ref. \onlinecite{renger2002relation} a secular and Markov approximation is carried out, yielding the correlation functions:
\begin{align}
 &\mathrm{tr}(\sigma_\alpha^-(t)B^\alpha_-(t) B^\alpha_+(0)\sigma_{\alpha}^+(0)\rho)\approx \nonumber\\
 &\qquad\mathrm{exp}(-i(E_\alpha-E_g-i\gamma_\alpha)t+g_\alpha(t)),
\end{align}
where $\gamma_\alpha$ contains the exciton-phonon scattering rates obtained in polaron picture Eq. \eqref{exciton_phonon_rate_polaron} and $g_\alpha(t)$ is the lineshape caused by diagonal exciton-phonon coupling:
\begin{align}
&g_\alpha(t)=\sum_{q \tilde{\alpha} G }\frac{|t_{\alpha\alpha}^{q G \tilde{\alpha}}|^2}{\hbar^2 \omega_{q\tilde\alpha}^2}(n_{q \tilde{\alpha}}(e^{i \omega_{q\tilde\alpha} t}-1) \nonumber\\&\qquad\qquad\qquad+(n_{q \tilde{\alpha}}+1)(e^{-i \omega_{q\tilde\alpha} t}-1)),
\end{align}
the typical form is known from the independent Boson model \cite{mahan2000many}.

\red{The following numerical evaluation of scattering rates, spectra, Green matrices, etc., always includes all 2500 calculated COM exciton states. No manual selection and truncation had been performed besides the restriction to the lowest 2500 states.}

\subsection{Analysis of scattering rates}

$\gamma_x$ is the inverse lifetime of the exciton states, plotted in Fig. \ref{plots_A} (d) for the $8\times4$ supercell over energy. The lifetime is mainly determined by exciton-phonon contribution, as the exciton-photon lifetimes all exceed nanoseconds or longer. 
For the localized states in areas A, B, and partly area C, the radiative lifetime is more or less the same as for states in the same energy area. For area C and higher, the radiative lifetime values show a broader range of values, but generally, the radiative lifetimes are higher since these states have less oscillator strength. Overall, the lifetime is dominated by exciton-phonon coupling, not radiative coupling for all states.

For quantum wells \cite{PhysRevB.95.235307,ZIMMERMANN200389,anderson1972size,rungezimmermannporterthomas}, few localized exciton states below the mobility edge of a single peak often have much stronger oscillator strengths and thus reduced radiative lifetimes, while the delocalized states from the same peak have increased radiative lifetimes. This is not the case for the heterolayer. In fact, for the heterolayer, many delocalized states have a shorter radiative lifetime than the localized states. (cf. Fig. \ref{plots_A} (d) ). 
One crucial difference is the position-dependent matrix element $\mathbf{D}(\mathbf{r})$ entering the exciton lifetime in Eq. \eqref{eq:exciton_photon_decay}, which is 1 for the quantum well and has its minimum at the position of the localized state (cf. Fig. \ref{dipolemomentplot} for dominating $\sigma_-$ polarization).
For most other states, the radiative lifetimes are more or less comparable low and differ from the quantum well case. Hence, the overall structure resembles more quantum dot states but with reduced optical activity inside a wetting layer as a fluctuating quantum well interface  (see discussion in Sec. \ref{sec:interlayer_exc_states}). 
The dominating exciton-phonon lifetimes for the standard Born-Markov case in Fig. \ref{plots_A} (d)  are very long for the first two bound states in areas A and partially B, so the overall lifetime in area A is dominated by radiative decay. However, for the exciton-phonon lifetimes in delocalized higher energy areas, more scattering channels are, even in Born-Markov approximation, open, and so for these states, exciton-phonon scattering dominates with their lifetime contribution ranging from microseconds to picoseconds.
In the Born-Markov approximation, all states in area A and some in area C have reduced exciton-phonon scattering rate, which is further reduced for lower temperatures. These lower energy states are dominated by phonon absorption and thus can only scatter to energetically higher states. 
In polaron transformation, exciton phonon scattering significantly contributes to the lower-lying states in areas A and B as the involved multi-phonon processes relax the dominance of one phonon absorption and result in a much lower temperature dependence. Overall, including the multiphoton processes, the lifetime distribution in Areas A and B are in the same order and longer compared to Areas C and D. There is a clear trend that lifetimes decrease towards lower energy. These deeper localized states have reduced scattering channels compared to the more delocalized states. Overall, this indicates that the so-called phonon bottleneck is entirely relaxed by including multi-phonons in polaron transformation. 
In Fig. \ref{plots_A} (d), the phonon lifetimes caused by acoustic and optical phonon scattering in with polaron transformation are plotted separately, as we see optical phonons have minimal impact on lifetimes as the lifetimes with and without optical phonons overlap. So, the lifetime of the excitons is predominantly determined by acoustic phonons. However, optical phonons are especially crucial for relaxation into the lower localized states, which happens on a longer time scale. Acoustic phonons cause the fast initial exciton reorganization, visible in the exciton relaxation Green's functions, which are analyzed later.

\begin{figure}[tb]
    \centering
    \includegraphics[width=8.6cm]{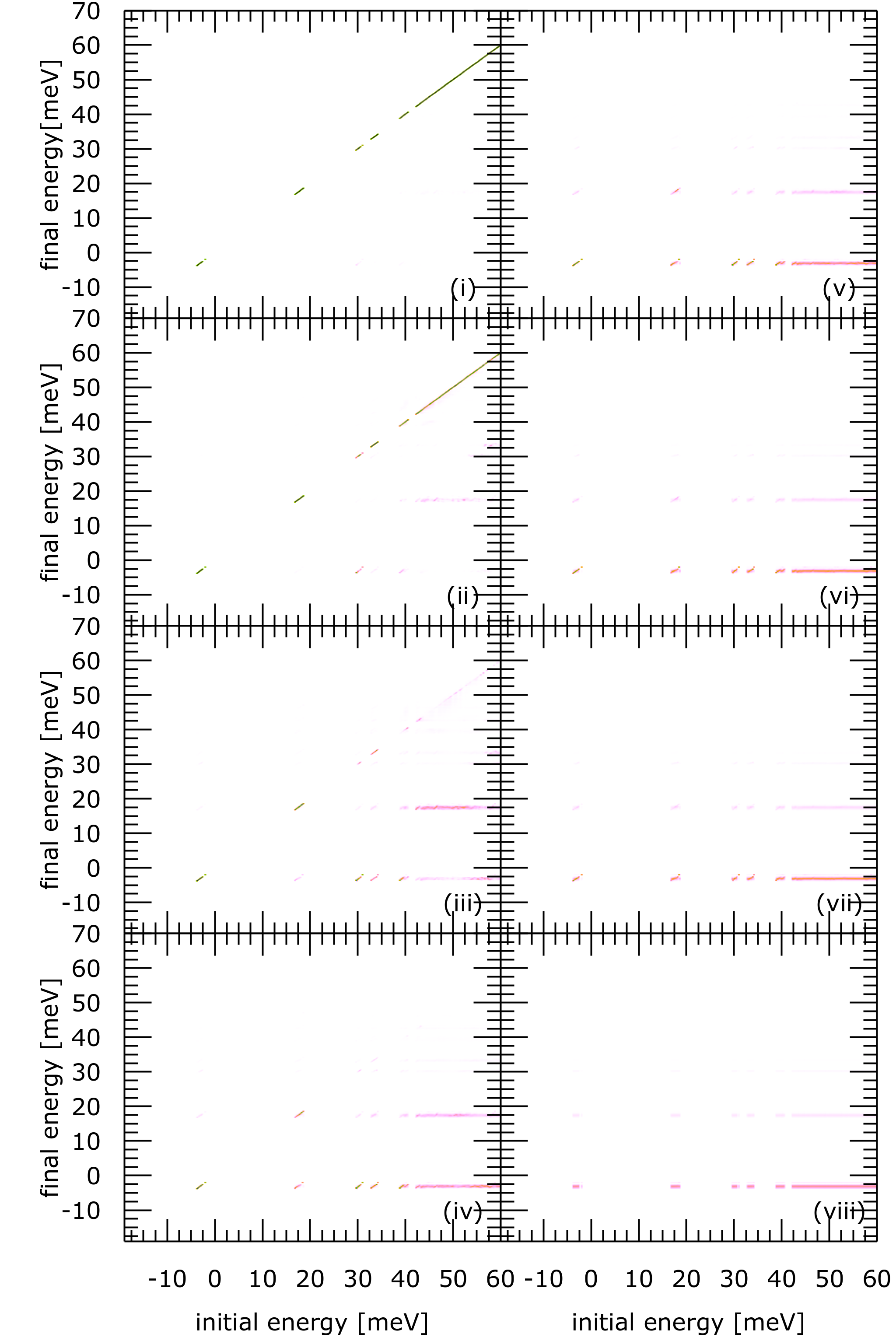}
    \caption{The exciton relaxation Green function WSe$_2$/MoSe$_2$ heterolayer in an 8x4 supercell using polaron theory at 5 K for a delay time between excitation and luminescence of (i) 100 fs, (ii) 1 ps, (iii) 10 ps, (iv) 40 ps, (v) 100 ps, (vi) 500 ps, (vii) 1 ns, (viii) 10 ns }
    \label{plots_G1}
\end{figure}

\begin{figure}[tb]
    \centering
    \includegraphics[width=8.6cm]{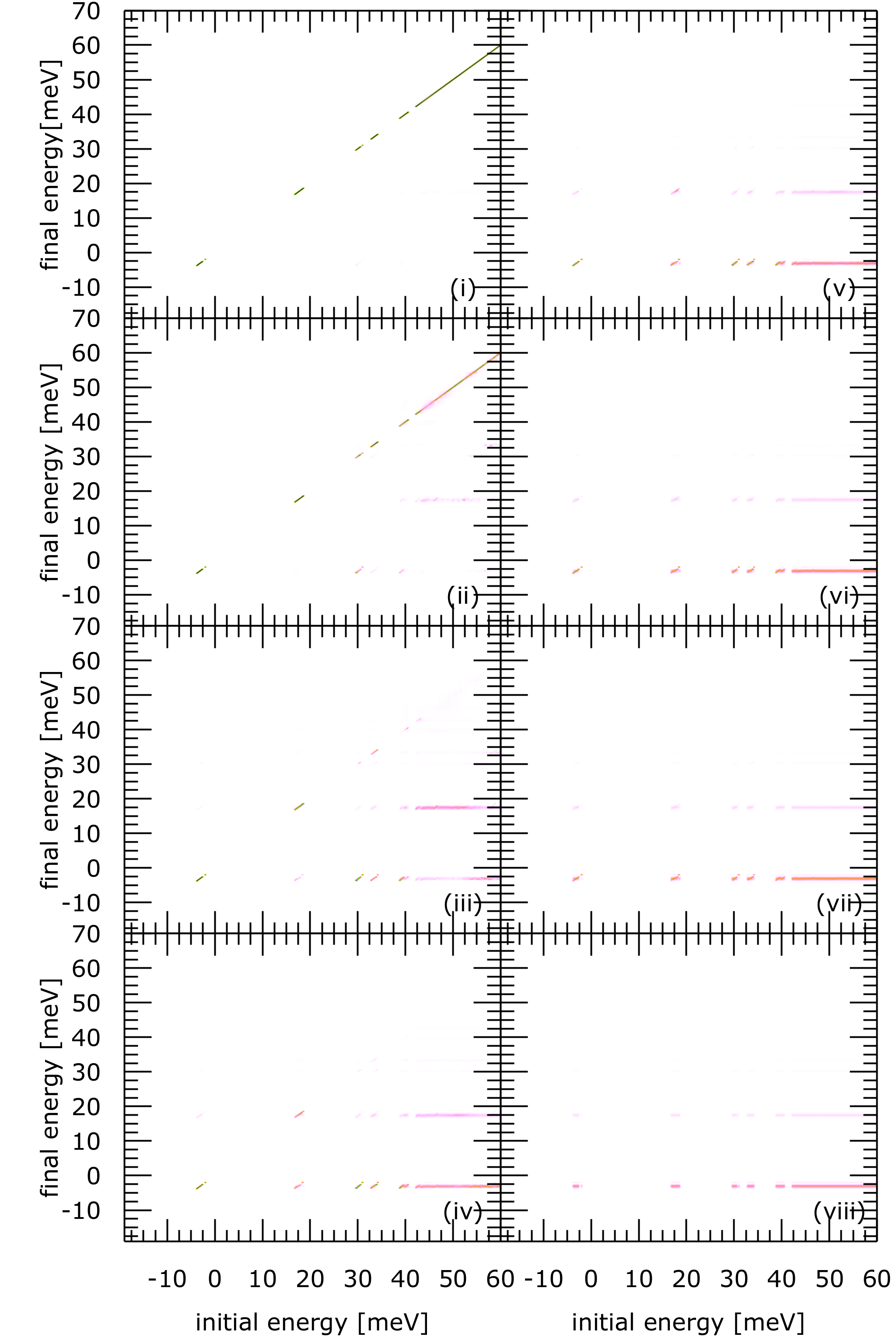}
    \caption{The exciton relaxation Green function WSe$_2$/MoSe$_2$ heterolayer in an 8x4 supercell using polaron theory at 20 K for a delay time between excitation and luminescence of (i) 100 fs, (ii) 1 ps, (iii) 10 ps, (iv) 40 ps, (v) 100 ps, (vi) 500 ps, (vii) 1 ns, (viii) 10 ns }
    \label{plots_G2}
\end{figure}
\subsection{Linear absorption spectrum}
A Fourier transform of the correlation function $\mathrm{tr}(\sigma_\alpha^-(t) \sigma_{\alpha}^+(0)\rho)$ yields the line shape $L_\alpha(\omega)$ of contribution from exciton $\alpha$ to linear absorption spectra. 
The contributions to linear absorption involve one exciton interaction with the optical field, including a coupling element. They are detected via the microscopic current, which includes another coupling element $\mathbf{D}_{\alpha g}=\int \mathrm{d}^2 r  \mathbf{D}(\mathbf{r}) \Psi_\alpha(\mathbf{r}) $.
Overall the linear absorption for incoming polarization $\mathbf{e}_p$ and detected polarization $\mathbf{e}_{p'}$ reads:
\begin{align}
    \alpha_{pp'}(\omega)=\sum_\alpha \mathbf{D}_{\alpha g}\cdot\mathbf{e}_{p} \mathbf{D}_{\alpha g}^*\cdot\mathbf{e}_{p}^* L_\alpha(\omega)
\end{align}
For the non-polaron case, we can retrieve the Lorentzian line shape to get:
\begin{align}
    \alpha_{pp'}^{np}(\omega)=\sum_\alpha \frac{\mathbf{D}_{\alpha g}\cdot\mathbf{e}_{p} \mathbf{D}_{\alpha g}^*\cdot\mathbf{e}_{p}^* \gamma_\alpha}{(E_\alpha-E_g-\omega)^2+\gamma_x^2}.
\end{align}

For the example of one random realization of the $8\times 4$ supercell, linear spectra are given in Fig. \ref{plots_C}.

In area A (up to area C), the spectra are dominated by many low-energy localized states inside the minima at different parts of the supercell. Most of the lowest localized states are well separated from other states and have, especially for low temperatures, a long life in Born-Markov approximation, resulting in many narrow resonances Fig. \ref{plots_C} a). Beginning in areas C and D close but below or above the mobility edges, the exciton lifetimes are much shorter (caused mainly by spontaneous emission to exciton states with lower energy) in the Born-Markov approximation, so it is challenging to locate the mobility edges even at low temperatures.
  The narrow resonances in areas A, B, and C in Fig. \ref{plots_C} a) have a line width far below typical optical resolutions in experiments.
For a plot comparable to experimental results, we convolute the calculated Born-Markov spectra with a Gaussian linewidth of 0.15 meV in Fig. \ref{plots_C} (a), which simulates typical experimental resolution.
Furthermore, the localized isolated states in areas A, B, and C are expected to have phonon side bands similar to quantum dots \cite{PhysRevLett.87.157401,krummheuer2002theory,forstner2003phonon}. The phonon sidebands are visible in the polaron calculation in Fig. \ref{plots_C} a) and dominate the localized states in areas A, B, and C. In contrast, in area D, the significant contribution of the line width seems to be the state's energy distribution and, thus, inhomogeneous broadening. As it is clear, for a correct physical picture of the states, the polaron picture is crucial, and the simple Born-Markov calculation (including convolution or no) will differ significantly from experiments. Therefore, in the following,  we will only include spectra calculated using the polaron in Fig. \ref{plots_C} b)-d).

In Fig. \ref{plots_C} (b), we study the polarization dependence of the different peaks.
Similar to the work without reconstruction, the peaks do not follow the expected valley selectivity of the single monolayer. The peaks in areas A and B have a relatively strict valley selectivity but with reversed polarizations compared to the monolayer case.
This is likely caused by the strong position dependence of the current element, which has a minimal $\sigma_-$ contribution in the potential minima (cf. Fig. \ref{dipolemomentplot}).
The two peaks in area C also show a strong valley selectivity, but the two peaks show an opposite valley selectivity. This property may be somewhat usable in optical devices such as transistors connecting or converting the two polarization directions.
The delocalized states in area D show almost no valley selectivity concerning polarization.

Fig. \ref{plots_C}(c) compares the linear absorption spectra for temperatures of $5K$, $20K$, and $50K$. For the localized states in areas A, B, and C, we see the strongest temperature dependence in the line shapes, as most of the homogeneous line shape is determined by phonons. However, the temperature dependence is not very strong as most of the line width is also caused by inhomogeneous broadening. Concerning the line shapes of the lower states, we have to note that the optical phonon satellite peaks, e.g., of areas B and C, overlap partially with the signatures in area A (the same is true for areas B and C) so that interpretation of the temperature dependence and dynamics of these peaks have to be carried out with caution.
 Only a minor effect is visible at increased temperatures for the delocalized states in area D (and at the mobility edge at the upper edge of area C). Here, most line shapes are determined by the distribution of the exciton states. Phonon line shapes for the delocalized states are also relatively small, as the nuclear reorganization for delocalized states is relatively small.

Comparing the linear spectra of two random realizations in Fig. \ref{plots_C}d) (and also with a $1\times 1$ supercell), we see no visible difference between the realizations in the 8x4 supercell (which is visible without polaron transformation in pure Born-Markov). Including polaron transformation, the 8x4 cell seems sufficient for describing the disorder in the structure. For the 1x1, a splitting of the states due to the phonon side peaks occurs, which is hidden in the 8x4 case inside the inhomogeneous broadening. 
 

In Fig. \ref{plots_C}d), a linear spectrum averaged over 20 realizations of the 8x4 supercell is shown (and two different contributing sets are shown) and shows no significant difference to the individual calculations.

\subsection{Exciton migration}
In nonlinear experiments, calculating the optical signals, including the full non-Markovian phonon dynamics, requires the computation of the multi-time correlation function. However, the nuclear wave packets can be neglected if the experiment involves sufficient long delay times longer than typical phonon reorganization times.
In this case, we can describe Exciton densities $\rho_{\alpha \alpha}$ dynamics using the calculated rates (non-polaron or polaron rates), respectively :
\begin{align}
\partial_t \rho_{\alpha\alpha}=-2\gamma_{\alpha}\rho_{\alpha \alpha}+\sum_{\tilde{\alpha}} \gamma_{\tilde{\alpha}\rightarrow \alpha }\rho_{\tilde{\alpha}\tilde{\alpha}},
\end{align}
which can also be written with a relaxation tensor $\Gamma_{\alpha\tilde{\alpha}}$ as:
\begin{align}
\partial_t \rho(t)=\sum_{\tilde{\alpha}} \Gamma_{\alpha'\tilde{\alpha}} \rho(t)
\end{align}
and define the relaxation Green's function $G_{\alpha\alpha'}(t_1,t)$ with $G_{\alpha\alpha'}(t_1,t_1)=\delta_{\alpha\alpha'}$ and \begin{align}
\partial_t G_{\alpha\alpha'}(t_1,t)=\sum_{\tilde{\alpha}}\Gamma_{\alpha'\tilde{\alpha}}G_{\alpha\tilde{\alpha}}(t_1,t)
\end{align}
The Green's function $G_{\alpha\alpha'}(\cdot,\cdot)$ tells us with which probability an exciton in state $\alpha$ at time $t_1$ ends up at exciton state $\alpha'$ at time $t$. Green's functions (including Green's function for the coherences) allow the calculation of various spectroscopic signals \cite{mukamel1995principles,abramavicius2009coherent}.
For plots over the energy of the Green's function, which we discuss later, the following form is used:
\begin{align}
G(E_f,E_i=\omega_{xg};T)=\sum_{\alpha'} G_{\alpha\alpha'}(T)\delta(E_f=\omega_{\alpha'g})
\end{align}
with the initial $E_i$ and final exciton energy $E_f$, respectively.

The relaxation-Green's function is an ingredient for the calculation of various spectroscopic signals such as photon echo, double quantum coherence \cite{abramavicius2009coherent,PhysRevB.95.235307,richter2018deconvolution}, pump-probe, or photoluminescence. In Fig. \ref{plots_G1} and \ref{plots_G2}, the relaxation Green's function is given for different delay times and temperatures ($T=5\mathrm{K}$ and $T=20\mathrm{K}$).
The initial relaxation occurs within the first tenth of picoseconds, after which almost all probability is concreted in areas A and B of localized states. 
The initial relaxation passes through many intermediate states, which can be seen as an off-diagonal contribution in Fig. \ref{plots_G1} and \ref{plots_G2}. After the relaxation, we see only contributions in the final states in areas A and B visible as a horizontal line at the final energy in the plot in Fig. \ref{plots_G1} and \ref{plots_G2}. 
The dynamics from 1 ns to 10 ns indicate that, finally, all excitons will be migrated to area A and that the lifetime of excitons in area B is in the nanosecond range.
Between $5\mathrm{K}$ and $20\mathrm{K}$, we see a faster relaxation time at higher temperatures and a higher exciton occupation at higher energy states in line with the equilibrium distribution.
The Green function calculated using polaron transformation does not show the phonon-bottleneck effect. A calculation in the Born-Markov calculation (not shown) leaves substantial occupation in areas C and D even for delay times in the nanosecond range due to the phonon bottleneck effect emerging in the lower order and quality of the Born-Markov calculation.

\section{Conclusion}
A framework was developed for simulating the interlayer exciton's dynamics, including exciton-photon scattering in transition metal dichalcogenides Moiré heterostructures suitable to include the influence of strain and address imperfections such as disorder and cracks. 
The framework was formulated in real rather than momentum space to include deviations from the standard translational periodic treatments.
For the influence of disorder, the localized exciton states show spectroscopic resonances broadened by the inhomogeneous exciton distribution.
After nanoseconds, only the lowest two exciton states are significantly occupied in polaron theory, including multi-phonon processes. In the Born-Markov approximation, the higher energy delocalized states would be occupied- a consequence of the phonon bottleneck effect caused by the single phonon approximation present in the Born-Markov approximation. 

The presented framework can be useful for calculating spectroscopic signals as it considers the influence of natural imperfections in experiments.

\begin{acknowledgements}
We thank Manuel Katzer and Andreas Knorr for stimulating discussions and introducing this problem field.
\end{acknowledgements}

\bibliography{strainbiblio}
\end{document}